\documentclass[12pt]{article}

\usepackage{amsmath}
\usepackage{amssymb}
\usepackage{fullpage}
\usepackage{xurl}
\usepackage{enumitem}
\usepackage[dvipsnames]{xcolor}

\usepackage{natbib}

\usepackage{todonotes}

\usepackage[normalem]{ulem}
\usepackage{cancel}

\usepackage[thref]{ntheorem}

\usepackage{enumerate}

\newcommand{\hS}{\widehat{S}}
\newcommand{\hC}{\widehat{C}}

\newcommand{\tS}{\widetilde{S}}

\newcommand{\hmu}{\widehat{\mu}}

\newcommand{\Hi}{{\mathcal{H}_i}}
\newcommand{\Ki}{{\mathcal{K}_i}}

\DeclareMathOperator{\Safe}{Safe}

\usepackage{tikz}

\usetikzlibrary{arrows}
\usetikzlibrary{shapes}
\usetikzlibrary{calc,fit}
\usetikzlibrary{patterns}
\usetikzlibrary{decorations.pathreplacing}

\usepackage{xr-hyper}
\usepackage[thref]{ntheorem}

\newtheorem{theorem}{Theorem}
\newtheorem{proposition}{Proposition}

\newtheorem{lemma}{Lemma}
\newtheorem{corollary}{Corollary}

\newtheorem{example}{Example}
\newtheorem{definition}{Definition}

\newenvironment{proof}{{\smallskip \noindent \bf Proof\quad}}{$\hfill\blacksquare$\bigskip}
\newenvironment{proofof}[1]{{\bigskip \noindent \bf Proof of Theorem \ref{#1}\quad}}{$\hfill\blacksquare$\bigskip}

\usepackage{xr-hyper}

\usepackage[colorlinks=true,allcolors=blue]{hyperref}
\usepackage{cleveref}

\begin{document}

\title{Incontestable Assignments\thanks{
We are especially indebted to Dhruva Bhaskar for his extraordinarily
extensive and detailed comments on successive drafts and for his
invaluable help in shaping and revising the manuscript.
We are grateful to Leeat Yariv and three
anonymous referees for their thoughtful comments and suggestions.  
We also thank Debraj Ray for extensive and valuable feedback on earlier versions. 
We are also grateful to 
    Dilip Abreu, In\'acio B\'o, Circe Haeringer Nadal, Qingmin Liu,
    Andrew Mackenzie, Vikram Manjunath,  
 Thayer Morrill, Michael Richter, Ariel Rubinstein, William Thomson,
 Alexander Westkamp, and 
  participants from the
  NYU theory seminar, the Ottawa Theory
  conference, and the 
  Matching and Market Design workshop at the WZB Berlin for their extensive comments and
  suggestions. 
  No external financial support was received for this research. The authors have no relevant financial relationships or other potential conflicts of interest to disclose.}
}

\author{Benoit Decerf\footnote{Universit\'e de Namur: benoit.decerf@unamur.be}\and
  Guillaume Haeringer\footnote{Baruch College: Guillaume.Haeringer@baruch.cuny.edu}
\and Martin Van der Linden\footnote{Independent scholar: martinvdl@gmail.com}}

\date{\today
}

\maketitle
\begin{abstract}
We study an information-based weakening of stability for school
choice, using the evidence students may invoke in appeals.
The model permits partial information about schools' choice
functions and the realized assignment. A student has a legitimate complaint if, in every
compatible state, she can block with some more-preferred school, and 
an assignment is incontestable if no student has such a complaint.
We characterize legitimate complaints using an ex-post safety-set.
For school-side information rules, under which the only admissible
assignment-side information is the student's own assignment and all
interim-compatible school-side states are preserved, incontestability
is equivalent to respecting interim safety-sets. 
Stable mechanisms and Top
Trading Cycles with responsive priorities are incontestable for
every such rule. Pareto improvements preserve this robustness.
When schools' choice functions are path-independent, all mechanisms that are
incontestable for every school-side information rule, are non-wasteful,
and guarantee group protection give each student the same attainable set
conditional on her preferences and interim information.

\end{abstract}

\section{Introduction}

Stability is a desirable objective when designing matching or assignment
mechanisms because it prevents eventual disruptions by blocking pairs, a
cause of market failure---see \cite{roth1991natural} and
\cite{kagel2000dynamics}.\footnote{It should be noted, though, that
stability does not always prevent market failure. See, for instance,
\cite{mckinney2005collapse}.}\textsuperscript{,}\footnote{Another common
justification for stability is fairness. However, in school choice
problems this normative criterion may come at the cost of efficiency
\citep{ergin2002efficient}.}
While this criterion is unquestionably needed in many cases, we contend
that stability may be stronger than necessary in settings like school
choice. In such markets, assignments are determined by a central
authority. Blocking pairs are not free to form independently, but arise
through the actions of individual students who can appeal their
assignment.

Appealing one's assignment differs from the standard blocking concept
because appeals generally need to be justified. A student may, for
instance, be able to establish from the admissions rules and the
information available to her that she should have received one of
several schools she prefers to her assignment. In contrast, a student
who is unhappy with her assignment but cannot provide a valid reason why
it should be different cannot substantiate a complaint. Students may
also not have the information needed to establish every blocking pair
that exists. They may have only partial information about the rule by
which schools select students, and privacy, disclosure, or evidentiary
rules may limit the information about other students that can be used
in an appeal. Thus, not every blocking pair is necessarily a threat to
an assignment. This is the point of departure for our paper: whether a
school choice assignment is appeal-proof depends on the information
available to students for substantiating an appeal.

The objective of this paper is to incorporate this information directly
into the stability concept. We allow students to have general partial
information about schools' choice functions and the realized assignment.
We interpret the information in the model as \textit{admissible
information}: information that a student can use to substantiate an
appeal, rather than necessarily everything she privately knows or
suspects. Depending on the setting, such information may be officially
provided by the school system or may be acquired by students in other
ways. We use ``information'' in this admissible sense throughout. We
also allow schools to have general choice functions, so responsive
priorities or substitutability are not needed for most of our results.

Given a student's admissible information, we consider all choice
profiles and assignments compatible with it. We say that she has a
\textit{legitimate complaint} if, at every such state, she can block
with some school she prefers to her assignment. The school need not be
the same in every compatible state. A student may therefore be able to
show that her assignment is defective without knowing which particular
school would sustain her complaint. An assignment is
\textit{incontestable} if no student has a legitimate complaint.

A simple example illustrates the idea. Consider four schools
$s_1,s_2,s_3,s_4$, each with one seat, and a student $i$ with preferences
\[
s_1\succ_i s_2\succ_i s_3\succ_i s_4\succ_i s_0.
\]
There are two other students, $j$ and $k$. Student $i$ knows that $j$ is
the only student with higher priority than her at both $s_1$ and $s_2$,
while $k$ is the only student with higher priority than her at both
$s_3$ and $s_4$. She does not know where $j$ and $k$ are assigned.
Since $j$ can occupy at most one seat, $i$ knows that she must be
accepted by at least one of $s_1$ and $s_2$. We call
$\{s_1,s_2\}$ a \textit{high-priority set}. Neither $\{s_1\}$ nor
$\{s_2\}$ is high-priority, since either school could be occupied by
$j$.

High-priority sets need not be nested. The upper contour sets generated
by the student's preferences are nested, however. We define the
student's \textit{safety-set} as the smallest upper contour set that is
high-priority. In the example, the safety-set is $\{s_1,s_2\}$.
Student $i$ cannot establish which of these schools would accept her,
but she can establish that at least one of them must. Hence, if she is
assigned to $s_3$, she has a legitimate complaint even though she cannot
identify a particular blocking school.

Our first result (Theorem~\ref{thm:new_theorem_1}) shows that this logic
is general. A student has a legitimate complaint if, and only if, her
assignment lies outside her ex-post safety-set. Equivalently, the least
preferred school in the safety-set gives a \textit{safety cutoff}: the
student has a legitimate complaint precisely when her assignment is
worse than this cutoff. Under complete information about schools'
choice functions and the realized assignment, the condition reduces to
the usual blocking condition, and incontestability coincides with
stability.

We then turn to assignment mechanisms. Since legitimate complaints
depend on students' admissible information, whether a mechanism is
incontestable also depends on the information environment. We capture
this through an \textit{information rule}, which specifies the
admissible information available to students at each problem and
assignment. Given a class of information rules $\Pi$, we say that a
mechanism is $\Pi$-incontestable if it generates no legitimate complaint
under any rule in the class. A class may capture variation in the
institutional rules governing admissible information or uncertainty
about what admissible information students acquire. More information
makes incontestability more demanding: a more informative rule
generates a weakly smaller safety-set. Consequently, if a class
contains a most informative rule, it is enough to check
incontestability under that rule.

We focus in particular on \textit{school-side information rules}. Under
these rules, students may have partial admissible information about
schools' choice functions but, once the assignment is realized, their
only admissible assignment-side information is their own assignment.
This seems a natural case for school choice: information about
admissions or priority rules may be available for an appeal while
information about other students' assignments may not be. For these
information rules, we can express the ex-post requirement of
incontestability in interim terms (i.e., before the assignment is
realized). We define an \textit{interim safety-set} using only the
student's school-side information. Although the interim and ex-post
safety-sets need not be the same,
Theorem~\ref{thm:equiv-interim-expost-top-set} shows that they give the
same answer to the relevant question: whether the student's realized
assignment belongs to the safety-set.

This result is particularly useful when we consider the class $\Pi^C$
containing all school-side information rules. This class has a most
informative rule, under which the full profile of schools' choice
functions is available as admissible information. A mechanism is
therefore $\Pi^C$-incontestable if, and only if, it always assigns each
student within her interim safety-set under full information about the
choice profile. We call this property \textit{respect for safety-sets}.

A key question, though, is what does this buy us? Crucially,
incontestability permits efficiency improvements that stability may rule
out. Stable mechanisms are $\Pi^C$-incontestable, but stability is not
necessary. Respect for safety-sets is preserved by Pareto improvements,
so any mechanism that Pareto dominates a $\Pi^C$-incontestable
mechanism is itself $\Pi^C$-incontestable. In particular, any mechanism
that Pareto improves a stable mechanism remains appeal-proof under every
school-side information rule. Under responsive priorities, Kesten's
\citeyearpar{kesten2010school} EADAM and Top Trading Cycles are also
$\Pi^C$-incontestable. TTC need not Pareto dominate a stable assignment,
so its inclusion does not simply follow from the Pareto-improvement
result. Incontestability is nevertheless not an ``anything goes''
requirement. The Boston mechanism can violate respect for safety-sets
and is therefore not $\Pi^C$-incontestable.

Finally, we ask what the interim safety-set tells us about the schools a
student can actually receive. Incontestability implies that it is an
upper bound on these schools. We show that, under substitutable and
consistent school choice functions and a weak additional condition on
the mechanism, this bound is exact.

The additional condition, which we call \textit{group protection}, has a
simple interpretation. Suppose a group of students can fill a collection
of schools, each student ranks her designated school first, and each
school would choose its designated subgroup over all students outside
the group. Group protection requires only that the group as a whole
retain the seats at these schools; it does not require the mechanism to
implement the particular first-choice assignment. Although this
condition gives us exactly the collective protection needed for the
attainability argument, it is weak. Stable mechanisms satisfy it on the
path-independent domain and, under responsive priorities, so do TTC,
EADAM, and even the Boston mechanism.

Our third main result (Theorem~\ref{thm:attainability}) shows that for
any $\Pi^C$-incontestable, non-wasteful mechanism satisfying group
protection, a student's set of attainable schools is exactly her interim
safety-set. With her preferences and interim information held fixed, every
school in the safety-set is assigned to her at some compatible problem,
and no school outside the set can be. Thus, mechanisms satisfying these
conditions may produce different assignments at a particular problem,
but they give each student the same set of possible schools conditional
on her interim information.

This characterization also gives a simple incentive result. Reny
\citeyearpar{reny2022efficient} shows that truth-telling is maxmin
optimal under his priority-efficient mechanism. We extend this result to
the mechanisms covered by our attainability theorem. Under truthful
reporting, the student's worst possible assignment is the safety cutoff,
and no misreport can improve this worst-case outcome. In our setting,
the worst case may range not only over the preferences of the other
students but also over school choice profiles compatible with the
student's information.

 \bigskip

\noindent \textbf{Related literature}
\qquad
Our paper intersects with several branches of the matching and
assignment literature. The first studies mechanisms that Pareto improve
the student-optimal stable mechanism. A key contribution is EADAM
\citep{kesten2010school}. Related rationales are provided by
\cite{ehlers2020legal} and \cite{reny2022efficient}. 
\cite{dur2019school} propose a family of mechanisms that allow for
certain forms of justified envy. Our approach permits Pareto
improvements over stable assignments because 
the information admissible for substantiating an appeal may be
incomplete, even when students privately know more, rather than because
justified envy is tolerated per se.

The paper also relates to the literature on matching and assignment
under partial information. First, the use of information sets to model
partial information is not new. It appears in work on regret-free
truth-telling \citep{fernandez2018deferred,chen2024regret}, 
transparent mechanisms \citep{moller2022transparent}, and auditability
\citep{grigoryan2023theory,grigoryan2024auditability}, among others.
In those papers, as in ours,
agents' information determines a set of compatible scenarios. But
rather than asking whether a student regrets a submitted preference
list, or whether she can verify that an announced mechanism was used,
we ask whether she can justify an appeal. The informational environment
is also different. In \cite{fernandez2018deferred} and
\cite{chen2024regret}, schools have responsive priorities and the
priority structure is fully available to students; in
\cite{moller2022transparent}, agents have access to object priorities
in a one-to-one assignment problem. We allow general choice functions,
and partial information about choice functions, capacities, and
assignments. 

Our paper is also related to work on stability under incomplete
information. \cite{chakraborty2010two} study a setting where agents
infer the types of potential partners from the observed matching,
\cite{yenmez2013incentive} considers ex-ante and interim stability
conditions,\footnote{See also \cite{ehlers2015matching}.} and
\cite{liu2014stable} study stability when firms do not observe
workers' types. We share with this literature the idea that blocking
must be evaluated relative to agents' information, but our focus is on
one-sided appeals in school choice.

Finally, our results connect to the literature on obvious
manipulations. \cite{troyan2020obvious} introduce non-obvious
manipulability and identify, in school-choice environments with known
responsive priorities, the schools relevant for students' worst-case
reasoning. \cite{decerf2021manipulability} develop a related safe-set
formulation. In the special case of known responsive priorities, our
interim safety-set coincides with their safe-set. The difference is
that this literature uses the object to study incentives, whereas we
use it to characterize legitimate complaints and incontestability.

\bigskip

The paper is organized as follows. Section~ \ref{sec:the_model}
presents the school-choice model together with a definition of students' information. Section
\ref{sec:legitimate-complaint} defines and characterizes legitimate complaints.
Section~\ref{sec:incont-mech} considers incontestable
mechanisms. Section~\ref{sec:robust-school-side} 
studies incontestable mechanisms for school-side information rules. We
discuss other classes of information rules 
in Section~\ref{sec:other-class-inform} and conclude in
Section~\ref{sec:conclusion}. Proofs are in the 
Appendix, with additional results in the online Appendix.

\section{The model}
\label{sec:the_model}

\subsection{School choice problems}
\label{sec:scho-choice-probl}

We consider school choice problems with a finite set $I$ of students
and a finite 
set $S$ of schools. The set $S$ contains a school denoted by $s_0$, which
we call the null school. Each student $i$'s preferences over schools are given by a strict
linear order $\succ_i$ over the set of schools $S$, and $s\succ_i
s'$ means that student $i$ prefers $s$ to $s'$, with $s,s'\in S$. 
A school $s$ is
unacceptable if $s_0\succ_i s$ (and acceptable otherwise).\footnote{So, $s_0$
captures students' outside option (i.e., not being assigned, which
could be, for instance, home schooling or being assigned to a school outside
the problem).} Given a 
set of students $J\subseteq I$, let $\succ_J=(\succ_i)_{i\in J}$. If
$J=I$, the set of all students, we will generally omit the subscript
and write $\succ$ instead of $\succ_I$.
We denote by $\mathcal{P}_i$ the set of all preference relations for a student $i$. We assume that for each $i\in I$, $\mathcal{P}_i$  consists of all strict preference orderings over $S$. Let $\mathcal{P}$ denote the set of all preference profiles and $\mathcal{P}_{-i}=\Pi_{j\neq i}\mathcal{P}_j$. 

Each school $s$ is endowed with a choice function
$C_s:2^I\to 2^I$, where for any set of students $J\subseteq I$,
$C_s(J)$ denotes the set of students chosen by $s$ from $J$.
We assume that $C_s(J)\subseteq J$ for every $J\subseteq I$ and
that $C_s(\varnothing)=\varnothing$.
We define the capacity of
school $s$ by
\begin{equation*}
  q_s(C_s):=\max_{J\subseteq I}|C_s(J)|,
\end{equation*}
the maximum number of students it can choose, and assume that
$q_s(C_s)>0$. We write $q_s$ when the choice function is clear
from the context.
For any set $J\subseteq I$, let
$R_s(J)=J\backslash C_s(J)$ denote the set of students rejected
from $J$. Following the standard school-choice convention that
schools do not deem students unacceptable, we assume that no school
rejects students from a set whose size is weakly below its capacity:
for each school $s$ and each set $J\subseteq I$, if
$|J|\leq q_s(C_s)$, then $C_s(J)=J$. For the null school, we
assume that $C_{s_0}(J)=J$ for every $J\subseteq I$, so that
$q_{s_0}(C_{s_0})=|I|$.

We denote by $\mathcal C_s$ the set of admissible choice functions
of school $s$, and let $\mathcal C=\Pi_{s\in S}\mathcal C_s$
denote the set of all choice profiles. For a subset of schools
$\hS\subseteq S$, let
$\mathcal C_{\hS}=\Pi_{s\in\hS}\mathcal C_s$.
Capacities may vary across choice functions in $\mathcal C_s$.
Until Section~\ref{sec:incont-mech}, we impose no additional
restrictions on these choice functions.

Since each choice profile $C$ determines the capacity vector
$q(C)=(q_s(C_s))_{s\in S}$, a \textbf{school choice problem}, for the
given sets of students and schools, is fully specified by
$(\succ,C)$.

Our intuitions can be 
easily captured using \emph{responsive
  priorities}. We say that a 
school $s$ has responsive priorities if its choice function $C_s$ can
be obtained from a strict 
priority ordering of students $r_s$ (with the convention that $r_s(i)<r_s(j)$ means that $i$ has
higher priority than $j$) as follows. For any set of students
$J\subseteq I$, $C_s(J)$ is obtained by ranking students in $J$ according to $r_s$ and admitting the 
highest-priority students up to capacity $q_s$.

\medskip

\noindent\textbf{Assignments:}
An assignment is a mapping
$\mu:I\cup S\to 2^I\cup S$ such that for each student $i\in I$,
$\mu(i)\in S$, for each school $s\in S$, $\mu(s)\in 2^I$, and
for each student $i\in I$, $\mu(i)=s$ if, and only if,
$i\in\mu(s)$. Let $\mathcal M$ denote the set of all assignments.
Given a choice profile $C$, let
\begin{equation*}
\mathcal M(C):=\{\mu\in\mathcal M:
|\mu(s)|\leq q_s(C_s)\text{ for every }s\in S\}
\end{equation*}
\noindent denote the set of feasible assignments. At a problem $(\succ,C)$,
an assignment is required to belong to $\mathcal M(C)$.

For a given set of students $J\subseteq I$, we will abuse notation and
denote by $\mu(J)$ the set of schools to which students in
$J$ are assigned, that is, $\mu(J)=\cup_{i\in J} \{\mu(i)\}$.
Similarly, for any set $T\subseteq S$, 
we write $\mu(T)$ to denote the set of
students that are assigned to the set of schools $T$ under
assignment $\mu$, that is, $\mu(T)=\cup_{s\in T}\mu(s)$. A student $i$'s preferences $\succ_i$ over schools implicitly define a
weak preference relation over assignments, denoted, with some abuse of
notation, by $\succeq_i$. So, $\mu\succeq_i \mu'$
if, and only if, $\mu(i)\succ_i\mu'(i)$ or $\mu(i)=\mu'(i)$.

\medskip

\noindent \textbf{Stability:} Stability is a standard concept in the school choice literature and is
often presented as the conjunction of three requirements: individual
rationality, non-wastefulness, and absence of blocking pairs (or
``justified envy''). In our setting, these requirements can be written
compactly as the absence 
of a blocking \emph{student--school pair} defined through the schools'
choice functions. Fix a problem $(\succ,C)$ and an assignment
$\mu$. We say that a student--school pair 
$(i,s)$ \textbf{blocks} $\mu$ if 
\begin{equation}
  \label{eq:stability-def}
  s\succ_i \mu(i)
  \quad\text{and}\quad
  i\in C_s\big(\mu(s)\cup\{i\}\big).
\end{equation}
An assignment $\mu$ is \textbf{stable} for a problem $(\succ,C)$ if it
admits no blocking pair.\footnote{In many-to-one matching problems,
stability is often also required to satisfy school-side individual
rationality: $C_s(\mu(s))=\mu(s)$ for every $s\in S$. Under our
maintained assumption that no student is unacceptable to any school,
this condition is automatic. Indeed, feasibility of an assignment
implies $|\mu(s)|\leq q_s$ for every school $s\in S$; hence, by the
no-unacceptability assumption, $C_s(\mu(s))=\mu(s)$ for every
$s\in S$.}
Stability rules out those situations in which a
student prefers a school and the school would choose that student if she
were added to its current assignment. Some immediate implications are:
1) Since all students can be assigned to the
null school, this implies individual rationality of stable
assignments. 2) Stable assignments cannot exhibit justified envy. 3)
If all schools find all students acceptable, then stable allocations
must be non-wasteful.

\subsection{Student Information}
\label{sec:information-sets}

We now define the information available to students about schools' choice functions and the realized assignment. We interpret this as the information students have access to, and which is admissible in the appeals procedure to substantiate an appeal. We take this information as a primitive of the model.

Each student $i \in I$ has an interim information set $\mathcal K_i
\subset \mathcal{C}$, describing her available school-side information
and the school choice profiles not ruled out by that information. Given a realized assignment $\mu$, student $i$'s ex-post
information is
  \begin{equation*}
\mathcal H_i
=\{(C',\mu')\in\mathcal K_i\times\mathcal A_i\,:\,
\mu'\in\mathcal M(C')\},
\end{equation*}
\noindent where $\mathcal A_i \subseteq \mathcal M_i(\mu)
\equiv
\{\mu'\in\mathcal M:\mu'(i)=\mu(i)\}$, 
which captures that ex-post, $i$ always knows her own assignment. The set $\Hi$ describes the set of school choice profiles and assignments compatible with her interim and assignment-side information.
Each choice profile $C'$ determines the schools' capacities
and hence the set $\mathcal M(C')$ of feasible assignments.
Thus, school-side information also restricts the possible capacities,
and an assignment is compatible with a choice profile only if it is
feasible under that profile.
We assume  that the information is always compatible with the truth: if $C$ is the true school choice profile and $\mu$ the realized assignment, then $C \in \mathcal{K}_i$, and $(C,\mu) \in \mathcal{H}_i$.

We do not make any additional assumption regarding the structure of
a student's information. For instance, if two choice profiles $C$ and
$C'$  are compatible with a student's information, we do not make any
assumption about how $C$ compares to $C'$. Our methodology works for
any type of partial information satisfying the structure we just laid
out. 
School-side and assignment-side restrictions are specified separately;
their interaction through feasibility is captured by
$\mu'\in\mathcal M(C')$. We do not allow additional joint
restrictions correlating choice profiles and assignments.

A key concept in the paper is the set of problems and ex-post realized
assignments that are compatible with a student's information.
With our set-based definition, compatibility is simply set membership
(or membership in a projection when we suppress some components). We
say that a choice profile $C$ or a pair $(C,\mu)$ is \emph{compatible}
with student $i$'s information set $\mathcal{K}_i$ or $\Hi$ if it is not ruled out by that
information: that is, the choice profile or the choice together with
the assignment is in the relevant information set.

\section{Legitimate complaints under partial information}
\label{sec:legitimate-complaint}

The standard notion of a blocking pair is defined relative to a
complete-information problem: given the realized assignment and
schools' choice functions, one asks whether a student prefers some
school to her assignment and that school would choose her over its
current assignees. Under partial information, however, a student may
be unable to identify the blocking pairs that she could form. We
therefore require a complaint to be justified by the information
available to the student. Specifically, a student has a legitimate
complaint only if her information establishes that she can block with
some more-preferred school in every problem and assignment compatible
with her admissible information.

\begin{definition}\label{def:legitimate_complaint} 
Let $\mu$ be an assignment, and let $i$ be a student with ex-post
information set $\Hi$. 
Student $i$ has a \textbf{legitimate complaint at $\mu$} if, for every
$(\hC,\hmu)$ compatible with $\Hi$, there exists a school $s$ such that
$s\succ_i \mu(i)$ and
\begin{equation}\label{eq:legit_complaint_block}
i\in \hC_s(\hmu(s)\cup\{i\}).
\end{equation}
\end{definition}

Given a problem $(\succ,C)$ and information profile $\mathcal{H}$, an
assignment $\mu$ is \textbf{incontestable} if no student has a
legitimate complaint.

Under full 
information, the statement that $i$ has a legitimate complaint is
equivalent to the statement that $i$ blocks with some school. An
important feature of the definition of a legitimate complaint is the
order of quantifiers: ``\textit{for each} compatible problem
\textit{there exists} a school such that \dots.'' 
That is,  the school
with which the student can block ~---$s$
in Eq.~\eqref{eq:legit_complaint_block}---~need not be the same across
compatible choice profiles and assignments.
A student may thus be able to conclude that she can block,
while being unable to determine with which school she can block and,
in the case of justified envy, against which student.

A legitimate complaint only concerns the student's \textit{own} assignment.  Under
complete information this restriction is immaterial: whenever an assignment
has a blocking pair, the student in that pair can identify it.  Under partial
information, however, a student may be able to infer that some blocking pair
exists without being able to establish that she belongs to one (for instance, because the information required to establish her own blocking pair is not admissible).  Our 
definition preserves the individual-standing structure of
ordinary blocking pairs, rather than allowing students to challenge
assignments on the basis of global instability or third-party
claims. In our setting, this is a restriction, but one which is
strongly motivated by institutional realities. School systems do not
typically allow students to make complaints based on information about
other students---for instance, doing so might violate the privacy of
other students, as even verifying or rebutting such a complaint could
reveal protected information about others. Our definition therefore
treats a complaint as legitimate only when the student's admissible
information establishes that her own assignment cannot be justified.

\subsection{High-priority sets and safety cutoffs}
\label{sec:high-priority-safety-cutoffs}

The definition of a legitimate complaint uses the idea that a student
may know that some set of schools is safe for her: for any choice
profile and assignment compatible with her information, she cannot be
rejected by all schools in that set. In other words, at least one of
these schools must accept her.

We call such a set an (ex-post) $\mathcal{H}_i$-high-priority set.
Formally, a non-empty set $\widehat{S}\subseteq S$ is an
\textbf{ex-post $\mathcal{H}_i$-high-priority set} for student $i$ if,
for every $(\widehat{C},\widehat{\mu})$ compatible with
$\mathcal{H}_i$,
\begin{equation}
\label{eq:Hi_high_priority_block}
i\in \widehat{C}_s\bigl(\widehat{\mu}(s)\cup\{i\}\bigr)
\quad\mbox{for some }s\in\widehat{S}.
\end{equation}

The property is monotone: if $\widehat{S}$ is an
$\mathcal{H}_i$-high-priority set for $i$, then any superset of
$\widehat{S}$ is also an $\mathcal{H}_i$-high-priority set. In
particular, since $q_{s_0}\geq |I|$ and
$i\in C_{s_0}(J\cup\{i\})$ for every $J\subseteq I$, any set
containing the null school is high-priority. Also, since any assignment
$\widehat{\mu}$ compatible with $\mathcal{H}_i$ maintains student
$i$'s assignment, so that $\widehat{\mu}(i)=\mu(i)$,
$\{\mu(i)\}$ is always an $\mathcal{H}_i$-high-priority set.
Hence, the collection of $\mathcal{H}_i$-high-priority sets is
nonempty.\footnote{Under known responsive priorities, a set $T$ is high-priority precisely
when no assignment compatible with $\mathcal{H}_i$ can fill every
school $s\in T$ with $q_s$ students who have higher priority than $i$.
When those responsive priorities are known by student $i$,
it is therefore relatively easy to check whether a given set
$\widehat{S}$ is a high-priority set. One only has to find a maximum
assignment restricting attention to the students ranked above $i$,
under the constraint that the assignment is compatible with
$\mathcal{H}_i$. If it is not possible to fill the capacity of every
school in $\widehat{S}$, then $\widehat{S}$ is a high-priority set.}

However, a high-priority set by itself does not tell us which
assignments leave $i$ without a legitimate complaint. For that, we
need a notion of a ``best'' high priority set for $i$, such that $i$
cannot be assigned something worse than everything in that set. While
in general, high-priority sets need not be nested, for each $s$, the
upper contour sets of schools $i$ weakly prefers to $s$,  
\begin{equation*}
  U_i(s):= \{t \in S: t \succeq_i s \}\,,  
\end{equation*}
\noindent are nested. Moreover, the smallest upper contour set (in the sense of set-inclusion) that is also a high-priority set gives us precisely the notion of a best set. Student $i$'s \textit{ex-post safety-set} of schools is therefore
\begin{equation}
    \label{eq:def_safety-set}
    \Safe^E (\succ_i,\Hi):= \min_{\subseteq} \{U_i(s): s\in S, U_i(s) \text{ is an $\Hi$-high-priority set}\}\,,
\end{equation}
where the superscript $E$ stands for ex-post.
The safety-set is the best set of schools, from $i$'s perspective,
within which she can guarantee admission to at least one school. This
property refers to the set as a whole---the student is guaranteed
acceptance by some school in the set, not by each of its
members. Moreover, membership in the safety-set does not a priori
imply that a school is attainable under a particular assignment
mechanism.\footnote{Theorem~\ref{thm:attainability} will establish
  this under some conditions.} The set is well-defined, since upper
contour sets are nested and the upper contour set $U_i(s_0)$ is a
high-priority set. Since $\Hi$ is ex-post information, $i$ knows her
own assignment, so any school in the safety-set must be at least as
good as $\mu(i)$. The safety-set therefore suggests a simple test for
a legitimate complaint: whether $i$ was assigned a school in her
safety-set. If not, she must have a legitimate complaint since the
safety-set is an upper contour set and all schools better than the
worst school in her safety-set are included.

Finally, notice that we can represent this requirement with a single
safety-cutoff: since the safety-set is an upper contour set, this test
only requires checking if $i$'s assignment is worse than the least
preferred school in her safety-set. We define
\begin{equation*}
  \underline s^E(\succ_i,\Hi):= \min_{\succ_i} \Safe^E (\succ_i,\Hi)
\end{equation*}
as $i$'s \textit{ex-post safety cutoff}. This gives us the following result:

\begin{theorem}\label{thm:new_theorem_1}
Let $\mu$ be an assignment, and let $i$ be a student with ex-post
information $\mathcal{H}_i$. Then $i$ has a legitimate complaint at
$\mu$ given $\mathcal{H}_i$ if and only if
\begin{equation*}
  \mu(i) \notin \Safe^E(\succ_i,\Hi),
\end{equation*}
\noindent 
or equivalently, if
\begin{equation*}
\underline{s}^E(\succ_i,\mathcal{H}_i)\succ_i\mu(i)\,.  
\end{equation*}
\end{theorem}
The safety cutoff is the best threshold for which student $i$ can
establish, from her information, that some school weakly better
than that threshold would choose her. The school providing this
guarantee can vary across compatible choice profiles and assignments;
the cutoff school itself need not choose her for every compatible
choice profile and assignment.
The cutoff is a threshold that an assignment must meet to generate
no legitimate complaint, rather than an unconditional lower bound
on the realized assignment. By
Theorem~\ref{thm:new_theorem_1}, $i$ has a legitimate complaint
exactly when her assignment is worse than this threshold.

The result nests the standard blocking logic under full
information. If students have full information, the safety-set
characterization reduces to the standard blocking condition. If $i$
forms a blocking pair with some school $s \succ_i \mu(i)$, then
$\{s\}$ is a high-priority set, so the safety-set lies entirely above
$i$'s assignment and $i$ has a legitimate complaint. Conversely, if
the assignment is stable, no school preferred to $\mu(i)$ would accept
$i$, so no upper contour set lying strictly above $\mu(i)$ can be
high-priority. Hence $i$'s safety-set contains her assignment and she
has no legitimate complaint.  

Our characterization does not require students to be able to compute
their safety-sets or safety cutoffs. It considers all complaints that
could in principle be established from their admissible
information. If students cannot draw all the implications of that
information, they may be able to substantiate fewer complaints, but an
incontestable assignment remains incontestable.

\subsection{Example}
\label{sec:examples}

The example illustrates how a publicly known admissions rule can
leave students with incomplete admissible information about schools' choice
functions, and how that information determines an ex-post safety-set. 

Consider a large urban district with ten schools \(s_h\), \(h=1,\ldots,10\), each with 100 seats, and 1,000 students. Each school gives priority to students who live in its walk zone. Within the walk-zone and non-walk-zone groups, students are ranked according to a common district examination score. Thus, school \(s_h\) first ranks the students eligible for its walk-zone priority, in examination-score order, and then ranks all remaining students in examination-score order. All students are acceptable.

There are exactly 30 students in the walk zone of each school, and the ten walk zones are disjoint. The admissions rule, school capacities, aggregate walk-zone counts, and disjointness of the zones are public. The district observes and verifies every student's address and examination score. In an individual appeal, however, a student may use her own records and the published aggregates but may not introduce the address or score records of specific students.

Consider a student \(i\) who is not eligible for walk-zone priority at any school and who ranks 300th in the district examination, so that exactly 299 students have a higher score. Her preferences are
\begin{equation*}
s_1\succ_i s_2\succ_i\cdots\succ_i s_{10}\succ_i s_0.
\end{equation*}
Student \(i\) does not know which named students are walk-zone
eligible at each school, but $i$ knows each school's capacity. Hence, although she knows the admissions rule, her admissible information does not identify the responsive priority ordering—and therefore the choice function—over named students. For instance, whether a lower-scoring student \(j\) ranks above \(i\) at \(s_h\) depends on whether \(j\) is eligible for \(s_h\)'s walk-zone priority.

We now calculate $i$'s ex-post safety-set when $i$ is assigned to
$s_6$. Let $\mathcal K_i$ denote her school-side information described
above, and suppose that she has no assignment-side information beyond
her own assignment. Her ex-post information set is therefore
\begin{equation*}
\mathcal H_i=\{(C',\mu'):C'\in\mathcal K_i,\
\mu'\in\mathcal M(C'),\ \mu'(i)=s_6\}.
\end{equation*}

It is possible for the first four schools to reject \(i\). Consider a
compatible classification in which all 299 higher-scoring students
live outside the ten walk zones. Each of \(s_1,\ldots,s_4\) can then
admit its 30 walk-zone students together with 70 distinct
higher-scoring students. This uses $4\times70=280\leq 299$
higher-scoring students. By contrast, if the first five schools all
rejected \(i\), their disjoint rosters would have to contain at least
$5\times70=350>299$ distinct students with higher examination
scores. This is impossible in every choice profile and assignment
compatible with \(i\)'s information. Hence, 
\begin{equation*}
  \Safe^E(\succ_i,\mathcal H_i)
   =\{s_1,s_2,s_3,s_4,s_5\}\,.
\end{equation*}
The guarantee applies to the set as a whole: for every compatible
choice profile and assignment, $i$ can block with at least one of
these five schools.
Since $s_6$ is not in $i$'s ex-post safety-set, $i$ has a legitimate
complaint. If $i$ were assigned to any school $s_h$ with $h\leq 5$,
then the previous analysis could be used to deduce that $i$'s ex-post
safety-set would be $\{s_1,\dots, s_h\}$, implying that $i$ would not
have a legitimate complaint. 

\section{Incontestable Assignment Mechanisms}
\label{sec:incont-mech}

The previous section characterized legitimate complaints for a fixed
assignment, conditional on students' preferences, school choice
functions, and the admissible information available to
students. Theorem \ref{thm:new_theorem_1} shows that, given student $i$’s 
admissible information $\Hi$, the absence of a legitimate complaint
for an assignment is equivalent to assigning $i$ a school in her safety-set. In this section, we move from specific assignments to
matching mechanisms and their incontestability. A matching mechanism
specifies the assignment that will be selected for each  school choice
problem $(\succ,C)$. Formally, a mechanism is a map $\varphi:\mathcal
P\times\mathcal C\to\mathcal M$ such that
$\varphi(\succ,C)\in\mathcal M(C)$ for every problem $(\succ,C)$.

\subsection{Information Rules and Incontestability}
\label{sec:inform-rules-incont}

The information available to students may depend on the realized school
choice problem and, ex post, on the realized assignment. We capture this
through an \textit{information rule}, which specifies the admissible information
available to each student in every such situation.

\begin{definition}\label{def:information_rule}
An \textbf{information rule} $\pi$ specifies, for each student $i$,
an interim information set
$\pi_i^I(\succ_i,C)\subseteq\mathcal C$ for each
$(\succ_i,C)\in\mathcal P_i\times\mathcal C$, and an ex-post
information set
$\pi_i^E(\succ_i,C,\mu)\subseteq\mathcal C\times\mathcal M$
for each $(\succ_i,C)\in\mathcal P_i\times\mathcal C$ and each
$\mu\in\mathcal M(C)$. The information sets induced by $\pi$ are
assumed to satisfy the requirements described in
Section~\ref{sec:information-sets}. In particular, every
$(C',\mu')\in\pi_i^E(\succ_i,C,\mu)$ satisfies
$\mu'\in\mathcal M(C')$.
\end{definition}

There are two ways to interpret the information generated by an
information rule. Under the first interpretation, all admissible
information is information officially provided to students by the
school system, for example through published admissions rules,
priority information, or information released after the
assignment. Information that is not disclosed by the school system
cannot be used to substantiate an appeal.\footnote{ 
For instance, England's School Admissions Code requires admission 
authorities to publish their admission arrangements, including
admission numbers and oversubscription criteria 
(\url{https://assets.publishing.service.gov.uk/government/uploads/system/uploads/attachment_data/file/1001050/School_admissions_code_2021.pdf}). 
Similarly, California Education Code \S 35160.5 publicly restricts
schools' choice functions (
\url{https://leginfo.legislature.ca.gov/faces/codes_displaySection.xhtml?lawCode=EDC&sectionNum=35160.5.}).}
Privacy restrictions provide one reason why some information may not
be disclosed. Under the second interpretation, admissible information
need not come only from the school system. Students may acquire
admissible information in other ways, and this information can also be
used to substantiate an appeal.

We ask
when a mechanism is incontestable not only under a single information
rule, but with respect to an entire class of information
rules. Specifically, we require that a single mechanism remain
incontestable for every information rule in a class $\Pi$, which we
define as a set of information rules.

\begin{definition}[Incontestability]
  For any information rule $\pi$, a mechanism $\varphi$ is
  \textbf{$\pi$-incontestable} if, for every problem $(\succ,C)$ and
  every student $i$, student $i$ has no legitimate complaint at
  $\varphi(\succ,C)$ given the ex-post information set
  $\pi_i^E(\succ_i,C,\varphi(\succ,C))$.  

  For any class of information rules $\Pi$, a mechanism $\varphi$ is
  \textbf{$\Pi$-incontestable} if $\varphi$ is $\pi$-incontestable for
  all $\pi \in \Pi$.  
\end{definition}

Under the first interpretation of information rules, a class of information rules represents the different rules governing disclosure and admissibility in appeals that arise across different
districts or implementations which will all use the same mechanism. To
remain incontestable across all these uses, the mechanism must
therefore be incontestable with respect to every information rule
used. The designer may also not know the specifics of every
information rule used by different districts, and therefore look for a
mechanism that works across a large set of possible rules. Under the
second interpretation, the requirement that the mechanism is
incontestable with respect to a class of rules can represent
uncertainty on the part of the designer about what admissible
information students can acquire. In this case, the class captures the
information rules the designer considers possible, and
incontestability with respect to that class is a robust criterion that
guarantees realized assignments are incontestable under all of them.

We now provide a general characterization of incontestability under
any class of information rules. For a fixed information rule $\pi$,
$\pi$-incontestability requires the assignment-level criterion in
Theorem~\ref{thm:new_theorem_1} to hold for every assignment generated
by the mechanism, using the ex-post information induced by $\pi$ at
that problem and assignment. For a class $\Pi$, the same no-complaint
condition must hold for every information rule in the class.
  The
following results express these requirements as safety-set conditions and
are key to analyzing incontestability for various classes of information
rules.\footnote{We omit the proofs of Corollary~\ref{thm2} and
  Proposition~\ref{lem:most_informative_rule} below because they are
  straightforward applications of Theorem~\ref{thm:new_theorem_1}. 
}

\begin{corollary}\label{thm2}
Let $\Pi$ be a non-empty class of information rules. A mechanism
$\varphi$ is $\Pi$-incontestable if and only if, for every
$(\succ,C)$, every student $i$, and every information rule $\pi \in
\Pi$,
\begin{equation*}
  \varphi_i(\succ,C) \in \Safe^E (\succ_i,\pi_i^E(\succ_i,C,\varphi(\succ,C)))\,.
\end{equation*}
\end{corollary}
This is an immediate corollary of Theorem \ref{thm:new_theorem_1}, but
we state it on its own as it clarifies the role of incontestability
with respect to classes of information rules. The characterization
states that for a mechanism to be $\Pi$-incontestable, in every
realized problem, the assignment to each student must belong to her
ex-post safety-set under every information rule in the class.

The ordering of information rules induces a corresponding ordering of
safety-sets. If an ex-post information set $\mathcal H_i$ is more
informative than $\mathcal H_i'$, so that
$\mathcal H_i\subseteq\mathcal H_i'$, then
\begin{equation*}
\Safe^E (\succ_i,\mathcal H_i)
\subseteq
\Safe^E(\succ_i,\mathcal H_i')\,.  
\end{equation*}

Indeed, every high-priority set under the coarser information set
$\mathcal H_i'$ is also high-priority under the finer information set
$\mathcal H_i$. Thus, if a class of information rules contains a most
informative rule, the safety-set condition generated by that rule is the
most demanding one.

We can compare information rules by the informativeness of the
information sets they generate. We say that an information rule $\pi$
is weakly more informative than an information rule $\pi'$ ex post if,
for every student $i$, every $(\succ_i,C)\in\mathcal P_i\times\mathcal C$,
and every $\mu\in\mathcal M(C)$,
\begin{equation*}
\pi_i^E(\succ_i,C,\mu)\subseteq\pi_i^{\prime E}(\succ_i,C,\mu).
\end{equation*}

Thus, every choice profile and assignment pair compatible with $\pi$
is also compatible with $\pi'$. This immediately gives us the
following result, which says that if a class $\Pi$ of information
rules contains a ``most informative'' information rule,
$\Pi$-incontestability reduces to checking that the mechanism assigns
each student in the ex-post safety-set generated by that information rule.

\begin{proposition}\label{lem:most_informative_rule}
Let $\Pi$ be a non-empty class of information rules. Suppose there
exists $\overline{\pi}\in\Pi$ such that $\overline{\pi}$ is weakly more informative
than every $\pi\in\Pi$. Then a mechanism $\varphi$ is
$\Pi$-incontestable if and only if, for every $(\succ,C)$ and every
student $i$, 
\begin{equation*}
\varphi_i(\succ,C)\in
\Safe^E\big(\succ_i,\overline{\pi}_i^E(\succ_i,C,\varphi(\succ,C))\big)\,.
\end{equation*}
\end{proposition}

\subsection{School-Side Information Rules}
\label{sec:scho-side-inform}

We now focus on information rules under which students may have admissible information about school choice functions, while ex post their only admissible assignment-side information is their own assignment.  An information rule $\pi$ is a
\textbf{school-side information rule} if, for every student $i$, every
$(\succ_i,C)\in\mathcal P_i\times\mathcal C$, and every
$\mu\in\mathcal M(C)$, 
\begin{equation*}
\pi_i^E(\succ_i,C,\mu)
=\{(C',\mu')\in\pi_i^I(\succ_i,C)\times\mathcal M_i(\mu):
\mu'\in\mathcal M(C')\}.
\end{equation*}

\noindent These rules are natural in many applications: information
about school priorities and capacities may be admissible, while
information about other students' assignments is not.   
Under such rules, we provide a sharper characterization of
incontestability. Incontestability as defined is inherently an ex-post
notion, because students appeal based on knowing their own
assignment. By Corollary \ref{thm2}, to ensure $\pi$-incontestability
for an information rule, we need to check that each student's
assignment belongs to her ex-post safety-set given her school-side and
assignment-side information.  As students know their own assignment,
this requires computing each ex-post safety-set for every possible
mechanism outcome. We will show that this can be reduced to checking a
single interim set for each student. We define analogous notions of
\textit{interim safety-sets} and \textit{interim safety-cutoffs},
which depend only on a student's school-side information. 

\begin{definition}
    Let $\mathcal K_i$ be an interim information set for student
    $i$. A nonempty set of schools $\hat S \subseteq S$ is an
    \textbf{interim $\mathcal{K}_i$-high-priority set} for student $i$
        if for every $C\in\mathcal K_i$ and every $\mu\in\mathcal M(C)$,
    \begin{equation*}
      i \in C_s(\mu(s)\cup \{i\}) \text{ for some } s\in \hat S.
    \end{equation*}
    Student $i$'s \textbf{interim safety-set} is
    \begin{equation*}
       \Safe^I (\succ_i,\Ki):= \min_{\subseteq}\{U_i(s): s\in S, U_i(s) \text{ is an interim $\Ki$-high-priority set}\}.
    \end{equation*}
    Student $i$'s \textbf{interim safety-cutoff} is
    \begin{equation*}
      \underline s^I(\succ_i,\Ki):= \min_{\succ_i} \Safe^I (\succ_i,\Ki).
    \end{equation*}
\end{definition}\footnotetext{It is easy to check that interim
  safety-sets satisfy the same monotonicity property as ex-post safety-sets:
  if 
$\mathcal K_i\subseteq\mathcal K_i'$, then
$\Safe^I(\succ_i,\mathcal K_i)
\subseteq
\Safe^I(\succ_i,\mathcal K_i')$.}

Interim safety-sets are conceptually similar to ex-post safety-sets,
but are computed before an assignment is realized and therefore range
over every compatible choice profile and every assignment feasible
under that profile.
The following example illustrates
how those two sets differ. 

\begin{example}
Consider two students, $i$ and $j$, and two schools, $s$ and $s'$,
each with one seat. Suppose that school $s$ ranks $j$ above $i$,
whereas school $s'$ ranks $i$ above $j$, and that student $i$ knows
this. Let $i$'s preferences be $s\succ_i s'\succ_i s_0$. Clearly, $\Safe^I(\succ_i,\{C\})=\{s,s'\}$.

Suppose now that $\mu(i)=s$, and that $\Hi$ is such that $i$ has no admissible information about $j$'s assignment. In the ex-post problem, the high-priority condition is
imposed only on assignments compatible with $i$'s admissible assignment-side information. Since $i$ already knows that she is assigned to
$s$, the singleton $\{s\}$ is an ex-post high-priority set for $i$.
Hence, $\Safe^E(\succ_i,\Hi)=\{s\}$.
\end{example}

Although the interim and ex-post safety-sets may differ, the following
result shows that for school-side information rules, they agree on the
relevant membership question: checking that the mechanism assigns
students to their ex-post safety-sets is equivalent to checking that
it assigns students to their interim safety-sets.  

\begin{theorem}\label{thm:equiv-interim-expost-top-set}
    Let $\Pi$ be a class of school-side information rules. For any
    $\pi \in \Pi$, problem $(\succ,C)$, assignment $\mu\in\mathcal{M}(C)$ and student
    $i$,
    \begin{equation*}
   \mu(i) \in \Safe^E(\succ_i,\pi^E_i(\succ_i,C,\mu)) \iff \mu(i) \in \Safe^I(\succ_i, \pi^I_i(\succ_i,C))\,.
 \end{equation*}
    Consequently, a mechanism $\varphi$ is $\Pi$-incontestable if and only if for every $\pi \in \Pi$, every problem $(\succ,C)$, and every student $i$, 
    \begin{equation*}
    \varphi_i(\succ,C) \in \Safe^I(\succ_i,\pi_i^I(\succ_i,C))\,,
  \end{equation*}
  or, equivalently, 
  $$ \varphi_i(\succ,C) \succeq_i \underline s^I(\succ_i, \pi^I_i(\succ_i,C)).$$
\end{theorem}

The equivalence concerns whether the realized assignment belongs to
the two safety-sets, not whether the safety-sets themselves are equal.
One direction is immediate. Under a school-side information rule, the
ex-post state space is
$\{(C',\mu'):C'\in\mathcal K_i,\
\mu'\in\mathcal M(C'),\ \mu'(i)=\mu(i)\},$
which is contained in the interim state space
$\{(C',\mu'):C'\in\mathcal K_i,\ \mu'\in\mathcal M(C')\}$.
Hence, ex-post information can only shrink the student's safety-set,
so membership in the ex-post safety-set implies membership in the
interim safety-set.

The intuition for the converse is that under a school-side information
rule, the only additional admissible assignment-side information ex
post is the student's own assignment. Conditioning on $\mu(i)$ may
therefore shrink the interim safety-set by 
effectively truncating it at $\mu(i)$: the student now knows that she 
has obtained a school at least as good as $\mu(i)$. But this
information cannot make the safety-set shrink past $\mu(i)$, since the
student has no additional admissible information about the assignment
at schools she prefers to 
$\mu(i)$. Thus, although the interim and ex-post safety-sets may
differ, learning one's own assignment cannot change the membership
verdict: the realized assignment belongs to the ex-post safety-set if
and only if it belongs to the interim safety-set.

As with the ex-post safety-set characterization, if a class has a most
informative rule (weakly more informative than any other rule in the
class), incontestability can be checked with respect to only that rule:

\begin{corollary}\label{cor:Pi-incontestable-top-sets}
    Let $\Pi$ be a class of school-side information rules that
    contains a most informative information rule $\bar \pi$. Then a
    mechanism $\varphi$ is $\Pi$-incontestable if and only if for
    every problem $(\succ,C)$ and every student $i$,
    \begin{equation*}
       \varphi_i(\succ,C) \in \Safe^I(\succ_i,\bar \pi_i^I(\succ_i,C))\,.
    \end{equation*}
\end{corollary}

\section{A Robust School-Side Information Class: $\Pi^C$}
\label{sec:robust-school-side}

We now consider a focal class of school-side information rules,
$\Pi^C$, which contains \textit{all} school-side information
rules. $\Pi^C$-incontestability takes a robust 
approach, requiring a
mechanism to be incontestable under \textit{every} possible
school-side information rule. Under the first interpretation of
information rules, this allows the disclosure and admissibility of
school-side information to vary arbitrarily. Under the second, it
allows the designer to be completely uncertain about what admissible
school-side information students can acquire.  

By the preceding theorem, this is
equivalent to requiring that the mechanism always assigns students in
the interim safety-set 
generated by every school-side rule. $\Pi^C$ contains a most
informative rule---namely the one under which the realized choice
profile $C$ is fully available as admissible information.  Corollary
\ref{cor:Pi-incontestable-top-sets} therefore implies 
that $\Pi^C$-incontestability can be checked using only the interim
safety-sets generated under full knowledge of $C$, which we denote
$\Safe^I(\succ_i,\{C\})$. To simplify terminology, we will say that a
mechanism \textbf{respects safety-sets} if, for every problem $(\succ,C)$
and every student $i$,
\begin{equation*}
  \varphi_i(\succ,C)\in \Safe^I(\succ_i,\{C\}).\footnotemark
\end{equation*}

Strictly speaking, this property is respect for
  interim safety sets; however, because it is evaluated under the most
  informative rule in \(\Pi^C\), which makes the realized choice
  profile available as admissible information, the relevant safety set
  is determined entirely by the 
  standard school choice problem \((\succ,C)\). We therefore omit
  “interim” and treat respect for safety sets as an ordinary property
  of an assignment at a school choice problem, and hence as a property
  of a mechanism.
  
When $C$ is induced by responsive priorities, the interim safety-set
$\Safe^I(\succ_i,\{C\})$ is precisely the safe set of
\cite{decerf2021manipulability}, and the least preferred school in
that set is the safety school of \cite{troyan2020obvious}, but
generalized to any school choice profile.  
Since $\Pi^C$ admits a most informative rule, a direct consequence of
Corollary \ref{cor:Pi-incontestable-top-sets} is the following: 

\begin{corollary}\label{cor:Pi-C_incontestable_iff_respect_top_sets}
A mechanism is $\Pi^C$-incontestable if and only if it respects
safety-sets.
\end{corollary}

\subsection{Incontestability and Efficiency}
\label{sec:incont-effic}

A priori, requiring incontestability under every school-side
information rule could be highly restrictive. The class $\Pi^C$
includes rules ranging from no admissible school-side information to full admissible information about the school choice profile, so a
$\Pi^C$-incontestable mechanism must remain appeal-proof regardless of
the precise school-side information available to students. 
The following proposition gives
the main applied implication of our characterization: several
prominent mechanisms that may produce unstable assignments~(including
mechanisms designed to improve efficiency relative to stable
assignments) remain incontestable with respect to $\Pi^C$.

\begin{proposition}\label{prop:family-incontestable-mech}
    The following mechanisms are $\Pi^C$-incontestable:
    \begin{itemize}
    \item[\textit{(a)}] Any stable mechanism.
    \item[\textit{(b)}] Under responsive priorities, Top Trading Cycles. 
    \item[\textit{(c)}] Under responsive priorities, Efficiency Adjusted Deferred Acceptance \citep{kesten2010school}.
    \item[\textit{(d)}] Any mechanism that Pareto-dominates a $\Pi^C$-incontestable mechanism. 
    \end{itemize}
  \end{proposition}

  The proof of Proposition~\ref{prop:family-incontestable-mech} can be
  found in
  Appendix~\ref{sec:proof_prop_family_incontestable}.\footnote{In the
    online appendix we show that two variations of the TTC mechanism
    proposed by \citep{morrill2015two} ---\textit{Clinch and Trade} and 
      \textit{First Clinch and Trade}--- are also $\Pi^C$-incontestable. }
    Statement \textit{(a)} is obvious (stable mechanisms are trivially
    incontestable and thus by
    Corollary~\ref{cor:Pi-C_incontestable_iff_respect_top_sets} respect
    safety-sets). Since the relevant interim safety-sets are upper contour sets, a Pareto improvement of an assignment that respects safety-sets must itself respect safety-sets. Hence \textit{(d)} follows. Since the Efficiency Adjusted 
Deferred Acceptance Pareto dominates any stable assignment
\citep[Theorem 1]{kesten2010school},
\textit{(a)} and \textit{(d)} together imply \textit{(c)}. 
  
As stability is a stronger notion than incontestability, of course any
stable mechanism is $\Pi^C$-incontestable. The main insight of the
result is that several mechanisms that are prominent and may produce
efficiency improvements are incontestable, despite violating
stability. In particular, the fourth point implies that any mechanism
that Pareto dominates a stable mechanism is
$\Pi^C$-incontestable. Therefore, if we think that students in
practice may not have access to or be permitted to use assignment-side
information in appeals, our result shows that these mechanisms are
still appeal-proof, and provides a rationale for their use.   

The proposition establishes robustness across the entire range of school-side information rules. Since $\Pi^C$
 contains every school-side information rule, any $\Pi^C$-incontestable mechanism is also incontestable with respect to every narrower class of such rules. The positive conclusions therefore do not depend on the precise school-side information available to students: the mechanisms remain incontestable whether students have very little admissible information about schools’ choice functions, full information, or anything in between. In this sense, $\Pi^C$-incontestability is the strongest test of appeal-proofness within the class of school-side information environments. By contrast, failure of $\Pi^C$-incontestability need not imply contestability under every narrower class, because coarser information may enlarge students’ interim safety-sets.

At the same time, the proposition is not an ``anything goes''
result. Efficiency alone is not sufficient for
$\Pi^C$-incontestability. As the following example shows, the Boston
mechanism may assign a student outside her interim safety-set, and is
therefore contestable under the most informative school-side
information rule, even though the resulting assignment is
efficient. Thus, while $\Pi^C$-incontestability permits important
departures from stability, it continues to impose a substantive
restriction on the assignments a mechanism may select.\footnote{We
  show in 
  the online Appendix that the Equitable TTC
  \citep{hakimov2018equitable} and the Application-Rejection mechanism
  \citep{chen2017chinese} do not respect  safety-sets.}

\begin{example}
Consider three students $i_h$, $h\in\{1,2,3\}$, and three schools
$s_h$, $h\in\{1,2,3\}$. Each school has one seat and schools have
responsive priorities. Preferences and priorities are depicted in
Table~\ref{tab:boston}.\footnote{All students rank $s_0$ last.}

\begin{table}[ht!]
  \centering
  \begin{tabular}{ccccccc}
    $i_1$ & $i_2$ & $i_3$ & \qquad & $s_1$ & $s_2$ & $s_3$\\
    \cline{1-3}\cline{5-7}
    $s_1$ & $s_1$ & $s_2$&   & $i_1$ & $i_1$&$i_1$\\
    $s_2$ & $s_2$& $s_1$&   & $i_2$& $i_2$&$i_2$\\
    $s_3$ & $s_3$& $s_3$&   & $i_3$& $i_3$&$i_3$\\
  \end{tabular}
  \caption{The Boston mechanism is not $\Pi^C$-incontestable}
  \label{tab:boston}
\end{table}

For student $i_2$, the set $\{s_1,s_2\}$ is a high-priority set at
$C$. Hence, $\Safe^I(\succ_{i_2},\{C\})=\{s_1,s_2\}$. Under the Boston
mechanism, however, given the preferences of students $i_1$ and
$i_3$, student $i_2$ is assigned to $s_3$. The assignment therefore
does not respect safety-sets, and thus the Boston mechanism is
not $\Pi^C$-incontestable.
\end{example}

\subsection{Attainability}
\label{sec:attainability2}

The preceding results show that a $\Pi^C$-incontestable mechanism must
assign each student within her interim safety-set. We now ask whether this
bound is exact: can every school in the interim safety-set arise as the
student's assignment at some problem compatible with her information?

For the remainder of this section, we assume that school choice functions
are substitutable and consistent. A choice function $C_s$ is
\textbf{consistent} if, for any $J,J'\subseteq I$,
\[
C_s(J)\subseteq J'\subseteq J
\quad\Longrightarrow\quad
C_s(J)=C_s(J'),
\]
and \textbf{substitutable} if, whenever $J\subseteq J'$,
\[
R_s(J)\subseteq R_s(J').
\]
These two conditions are equivalent to path independence
\citep{aizerman1981general,moulin1985choice}:
\[
C_s(J\cup J')=C_s(C_s(J)\cup C_s(J'))
\]
for every $J,J'\subseteq I$.

To see what is needed for attainability, fix a school $s^\ast$ in student $i$'s
interim safety-set, and let $\widehat S$ be the set of schools that $i$
strictly prefers to $s^\ast$. We want to construct a problem compatible with $i$'s interim information at which
$i$ is assigned to $s^\ast$. Since $s^\ast$ belongs to the safety-set,
$\widehat S$ cannot itself be an interim high-priority set: otherwise
there would be a smaller high-priority upper-contour set than the
safety-set. Hence there is a choice profile $C$ compatible with $i$'s
information and a feasible assignment under which $i$ is rejected by
every school in $\widehat S$. Path independence allows us to strengthen
this construction. We can fill the schools in $\widehat S$ with a group
of other students so that these schools would not replace members of the
group with students outside it, while $i$ remains outside the group.
Formally, there is a feasible assignment $\widehat\mu$ such that, writing
\[
G=\widehat\mu(\widehat S)
\qquad\text{and}\qquad
Z=I\setminus G,
\]
we have $i\in Z$ and
\[
\widehat\mu(s)=C_s(\widehat\mu(s)\cup Z)
\]
for every $s\in\widehat S$.

We can now choose the preferences of the other students so that every
student in $\widehat\mu(s)$ ranks $s$ first. If these students continue
to fill the schools in $\widehat S$ under the mechanism, then $i$ cannot
be assigned to any school she prefers to $s^\ast$. We can also make
$s^\ast$ unacceptable to every other student. Individual rationality and
non-wastefulness then imply that $i$ must be assigned to $s^\ast$.
Thus, the only additional requirement we need from the mechanism is that
the group $G$ continue to occupy the seats at the schools in
$\widehat S$. This motivates the following property.

\begin{definition}\label{def:group-protection2}
A mechanism $\varphi$ satisfies \textbf{group protection} if the following
holds at every problem $(\succ,C)$. Let
$T\subseteq S\setminus\{s_0\}$ be nonempty, and let
$(G_s)_{s\in T}$ be pairwise disjoint sets of students. Write
\[
G=\bigcup_{s\in T}G_s,
\qquad
Z=I\setminus G.
\]
Suppose that, for every $s\in T$,
\begin{itemize}
\item[\textit{(i)}] each student $j\in G_s$ ranks $s$ first;
\item[\textit{(ii)}] $|G_s|=q_s(C_s)$; and
\item[\textit{(iii)}] $C_s(G_s\cup Z)=G_s$.
\end{itemize}
Then
\[
\bigcup_{s\in T}\varphi_s(\succ,C)=G.
\]
\end{definition}

Group protection only requires that the group collectively retain the
seats at the schools in $T$. It does not require that each student in
$G_s$ actually be assigned to $s$. Since group protection provides exactly the collective protection used in the argument above, a natural question is how restrictive the condition is. It turns out to be quite weak. For instance, it is weaker than first-choice stability, which requires that no student have justified envy at her first-choice school. Moreover, the condition is satisfied by stable
mechanisms on the path-independent domain. Under responsive priorities,
TTC and Kesten's EADAM also satisfy group protection, as does the Boston
mechanism.\footnote{Proofs are available in the Online Appendix.}

We can therefore state the attainability result as follows.

\begin{theorem}\label{thm:attainability}
Suppose that, for every $C'\in\mathcal C$ and every $s\in S$,
$C'_s$ is path independent. Let $\varphi$ be a
$\Pi^C$-incontestable and non-wasteful mechanism satisfying group
protection. Fix an information rule $\pi\in\Pi^C$, a problem
$(\succ,C)$, and a student $i$, and let
$\mathcal K_i=\pi_i^I(\succ_i,C)$.
Then, for every school $s\in S$,
\[
s\in\Safe^I(\succ_i,\mathcal K_i)
\quad\Longleftrightarrow\quad
\varphi_i\bigl((\succ_i,\widehat\succ_{-i}),\widehat C\bigr)=s
\]
for some $\widehat C\in\mathcal K_i$ and some
$\widehat\succ_{-i}\in\mathcal P_{-i}$.
\end{theorem}

Thus, the interim safety-set is not only an upper bound on the student's
possible assignments. It is exactly the set of schools she can obtain
across problems compatible with her interim information.

An immediate implication is an attainable-set equivalence across
mechanisms. Any two $\Pi^C$-incontestable, non-wasteful mechanisms
satisfying 
group protection  generate exactly the same set of attainable
schools for every student and every interim information set: that set
is $\Safe^I(\succ_i,\mathcal K_i)$. The mechanisms may select different
assignments at a particular problem, but from the student's interim
perspective they offer the same range of possible schools.

This equivalence concerns attainable outcomes rather than the
pointwise behavior of the mechanisms. The mechanisms need not select
the same assignment at a given problem, nor need they have the same
incentive or efficiency properties. Rather, conditional on a
student's preferences and interim information, the identity of the
mechanism does not affect the set of schools that she may receive.
In this specific sense, the mechanisms are indistinguishable from the
student's interim perspective.\footnote{This conclusion is related in
spirit to the notion of $i$-indistinguishability in \citet{li2017obviously},
but is distinct from it. Li's notion concerns an agent's experience
within an extensive-form mechanism, whereas our result concerns only
the set of final schools attainable by the student.}

Theorem~\ref{thm:attainability} is also related to Lemma~1 and
Corollary~1 of \citet{troyan2020obvious}, who obtain a similar
attainability result for stable-dominating mechanisms when schools
have responsive priorities that are perfectly known to students.
The two results are not formally nested. Their result does not impose
group protection, whereas our theorem applies to mechanisms, such
as TTC, that need not Pareto dominate a stable mechanism. Our result
also allows for general school-side information and consistent and
substitutable choice functions. 

\subsection{Maxmin Incentives}
\label{sec:incentives}

We now use the characterization of attainable schools to derive a maxmin incentive property
for mechanisms satisfying the conditions of
Theorem~\ref{thm:attainability}. \cite{reny2022efficient} studies the incentive properties of the
Efficiency Adjusted Deferred Acceptance Mechanism
\citep{kesten2010school}.\footnote{\cite{reny2022efficient} proposes a
different mechanism than EADAM, but the two are outcome equivalent when
all students waive their priorities.}
That mechanism is not strategy-proof, but Reny shows that it still
satisfies a robust incentive property: if students compare strategies
by their worst-case outcome, according to their true preferences, then
truth-telling is optimal. This belief-free criterion is close in spirit
to our use of safety-sets. In our framework, a student may also be
uncertain about schools' choice functions, so the relevant worst case
ranges over all problems compatible with the student's interim
information.

Fix a mechanism $\varphi$ and a student $i$ with true preferences
$\succ_i$ and interim information set $\mathcal K_i\subseteq\mathcal C$.
When student $i$ submits a preference ordering
$\succ_i'\in\mathcal P_i$, let
$w_i(\succ_i'\mid\succ_i,\varphi,\mathcal K_i)$ denote the worst
school, according to $\succ_i$, that $i$ can receive as the choice
profile ranges over $\mathcal K_i$ and the preferences of the other
students range over $\mathcal P_{-i}$. Formally,
\begin{equation}
  \label{eq:worst_school}
  w_i(\succ_i'\mid\succ_i,\varphi,\mathcal K_i)
  \equiv
  \min_{\substack{C\in\mathcal K_i\\
                  \succ_{-i}\in\mathcal P_{-i}}}
  \varphi_i\bigl((\succ_i',\succ_{-i}),C\bigr),
\end{equation}
where the minimum is taken with respect to $\succ_i$.

Theorem~\ref{thm:attainability}  gives the
following extension of Reny's maxmin result.\footnote{The proof can be
  found in Appendix~\ref{sec:maxmin-incentives}.} 

\begin{proposition}\label{prop:maxmin-truthful}
Fix a student $i$ with interim information set $\mathcal K_i$. 
Let $\varphi$ be a
$\Pi^C$-incontestable and non-wasteful mechanism satisfying group
protection. 
Then, for every preference ordering $\succ_i'\in \mathcal P_i$,
\begin{equation}
  \label{eq:maxmin_truthful}
  w_i(\succ_i|\succ_i,\varphi,\mathcal K_i)\ \succeq_i\
  w_i(\succ_i'|\succ_i,\varphi,\mathcal K_i).
\end{equation}
\end{proposition}

The intuition follows directly from the attainability result. Under
truthful reporting, the set of schools that student $i$ can receive is
exactly $\Safe^I(\succ_i,\mathcal K_i)$ with safety-cutoff $\underline
s^I(\succ_i,\Ki)$, which is therefore the student's worst-case outcome
under truthful reporting. Under any alternative report 
$\succ_i'$, the attainable set is
$\Safe^I(\succ_i',\mathcal K_i)$. This set is itself an interim
$\mathcal K_i$-high-priority set, but it is selected using the reported 
rather than the true preferences. Let $s$ be the least-preferred
school in $\Safe^I(\succ_i',\Ki)$ according to the true preference
$\succ_i$, so this is the worst-case outcome under the misreport. Then
$U_i(s)$ contains $\Safe^I(\succ_i',\Ki)$, and is therefore also
high-priority. Since $\Safe^I(\succ_i,\Ki)$ is the smallest
high-priority upper contour set according to the true preferences, its
safety-cutoff must be weakly preferred to $s$. A misreport may
therefore improve the student's assignment at some problems, but it
cannot improve her worst-case outcome.

\section{Other classes of information rules}
\label{sec:other-class-inform}

To place $\Pi^C$ in context, consider two boundary classes.
For the assignment-side comparison, assume that $\mathcal C_s$
contains every responsive choice function with capacity in
$\{1,\ldots,|I|\}$ at each real school $s$; other admissible choice
functions may also be included. Capacities are determined by the
choice functions, as in Section~\ref{sec:scho-choice-probl}.
The class $\Pi^\mu$ consists of all rules with
$\pi_i^I(\succ_i,C)=\mathcal C$ and
\begin{equation*}
\pi_i^E(\succ_i,C,\mu)
=\{(C',\mu')\in\mathcal C\times\mathcal A_i:
\mu'\in\mathcal M(C')\},
\end{equation*}
where $\mu\in\mathcal A_i\subseteq\mathcal M_i(\mu)$.
Thus admissible information may restrict the compatible assignments,
but imposes no restrictions on choice rules or capacities beyond
the maintained model restrictions and compatibility with those
assignments.

A mechanism is $\Pi^\mu$-incontestable if and only if it is
individually rational and never leaves a preferred real school empty:
for every $(\succ,C)$, every student $i$, and every $s\neq s_0$,
\begin{equation*}
s\succ_i\varphi_i(\succ,C)
\quad\Longrightarrow\quad
\varphi_s(\succ,C)\neq\varnothing.
\end{equation*}
When the entire realized assignment is admissible information,
an empty preferred school yields a complaint because every
compatible capacity is positive.

Conversely, if the assignment is individually rational and every
preferred real school is nonempty, each such school's choice rule can
be taken, consistently with the admissible information, to be
responsive, with capacity equal to its enrollment and its current
assignees ranked above the complaining student. 
This gives a compatible state in which that student cannot block.

The class $\Pi^{C,\mu}$ consists of all admissible information rules
about choice profiles and assignments.
It contains the rule under which the entire realized choice profile
and assignment are admissible information, so that a legitimate
complaint is an ordinary blocking complaint.
Consequently, $\Pi^{C,\mu}$-incontestability is equivalent
to stability. Necessity follows by considering this rule.
Sufficiency
follows because the true stable choice-profile/assignment pair is
compatible with every student's information.

These boundary cases clarify the role of $\Pi^C$.
Incontestability under every rule in $\Pi^\mu$ rules out
unacceptable assignments and empty preferred schools;
incontestability under every rule in $\Pi^{C,\mu}$ imposes
stability.
School-side information can support claims even when the only
admissible assignment-side information is the student's own
assignment, while still permitting departures from stability.
Formal proofs are in the Online Appendix.

\section{Conclusion}
\label{sec:conclusion}

We have studied incontestability as a weakening of stability that
takes account of the information students can use to substantiate an
appeal. Rather than assuming a particular information structure, we
allow general partial information about schools' choice functions and
the realized assignment. Our safety-set characterization tells us when
a student can establish a legitimate complaint given her admissible
information. We then use information rules to study which mechanisms
remain incontestable under different information environments. 

Our main results focus on school-side information rules, under which
students may have information about schools' choice functions, but
their only admissible assignment-side information is their own
assignment. For these rules, incontestability can be characterized
using interim safety-sets. Importantly, stability is not necessary:
incontestability is preserved by Pareto improvements, so mechanisms
can improve efficiency relative to stable assignments while remaining
incontestable. At the same time, incontestability is not an ``anything
goes'' requirement, as the Boston mechanism need not satisfy
it. Figure 1 summarizes how several familiar mechanisms compare. We
also show that, under path independence and group protection,
mechanisms that are $\Pi^C$-incontestable and non-wasteful give each
student the same set of attainable schools conditional on her
preferences and interim information, even though they may produce
different assignments at a particular problem. 

Our approach opens up several potential directions for future research
on how information affects the design of assignment mechanisms. One
direction is to study particular institutional rules governing
admissible information and their implications for
incontestability. Another is to extend the analysis to two-sided
matching environments, where both sides may have information that
allows them to jointly challenge an assignment.
A related open question relates to information sharing. An assignment may be
incontestable, yet
become contestable once students pool their information and submit a joint complaint.
Characterizing information rules that are joint-complaint-proof would clarify when
information sharing restores the force of stability.

More broadly, we believe that
incorporating the information agents can use to substantiate
complaints into the analysis of stability offers scope for further
work in the design of matching mechanisms.

\begin{figure}[t!]
	\centering
	\begin{tikzpicture}[scale=1]
		
		\draw[red,rounded corners,line width=2pt] (-0.7,-0.5) rectangle (3.5 ,2.8) {};
		\node[red] at (2.6,-0.2) {\Large\textsf{Stable}};
		
		\draw[blue,rounded corners,line width=2pt] (-1.4,0.3) rectangle (11.8,3) {};
		\node[blue] at (10,2.6) {\Large\textsf{Strategyproof}};
		
		\draw[teal,rounded corners,line width=2pt] (4.2,0.5) rectangle (12.1,5.2) {};
		\node[teal] at (11,4.9) {\Large\textsf{Efficient}};
		
		\draw[Bittersweet,dashed,rounded corners,line width=2pt] (-0.9,1.2) -- (-1,6) -- (7.7,6)
		-- (7.7,3.2) -- (3.8,3.2) -- (3.8, 1.2) -- (-0.9, 1.2) {};
		\node[Bittersweet] at (2.5,5.7) {\large\textsf{Weakly Pareto Dominates SOSM}};

		\draw[Purple ,rounded corners,line width=5pt] (-1.2,-1) rectangle (8,7.2) {};
		\node[Purple] at (1.4,6.8) {\Large$\mathbf{\Pi^C}$\textsf{\textbf{-Incontestable}}};

		\node[] at (1.4,2) {\Large\textsf{SOSM}};
		
		\node[] at (6,2) {\Large\textsf{TTC}};
		
		\node[] at (5,1) {\Large\textsf{CT}};
		
		\node[] at (7,1) {\Large\textsf{FCT}};
		
		\node[] at (10,1) {\Large\textsf{ETTC}};
		
		\node[] at (10,4) {\Large\textsf{Boston}};
		
		\node[] at (6,4.3) {\Large\textsf{EADAM}};
		\node[] at (6,3.7) {\Large\textsf{\textit{(full consent)}}};
		
		\node[] at (1.4,4.4) {\Large\textsf{EADAM}};
		\node[] at (1.4,3.8) {\Large\textsf{\textit{(partial consent)}}};
		
		\node[] at (11.7,6.8) {\Large\textsf{Application-Rejection}};
		\node[] at (11.7,6.2) {\Large $(1<e<|S|)$ };

	\end{tikzpicture}
	\caption{Comparison of mechanisms}
	\label{fig_Venn}
	\bigskip
	\begin{minipage}{1\textwidth}
		\footnotesize{
				Note: CT denotes ``Clinch and Trade'',
                                FCT denotes ``First Clinch and
                                Trade'', ETTC denotes ``Equitable
                                TTC''. We show in the Online Appendix
                                that CT and FCT respect safety-sets, while ETTC
                                violates respect for  safety-sets. A
                                description of CT, FCT,
                                Application-Rejection, and ETTC can be
                                found in the online Appendix.
		}
	\end{minipage}	
	
\end{figure}
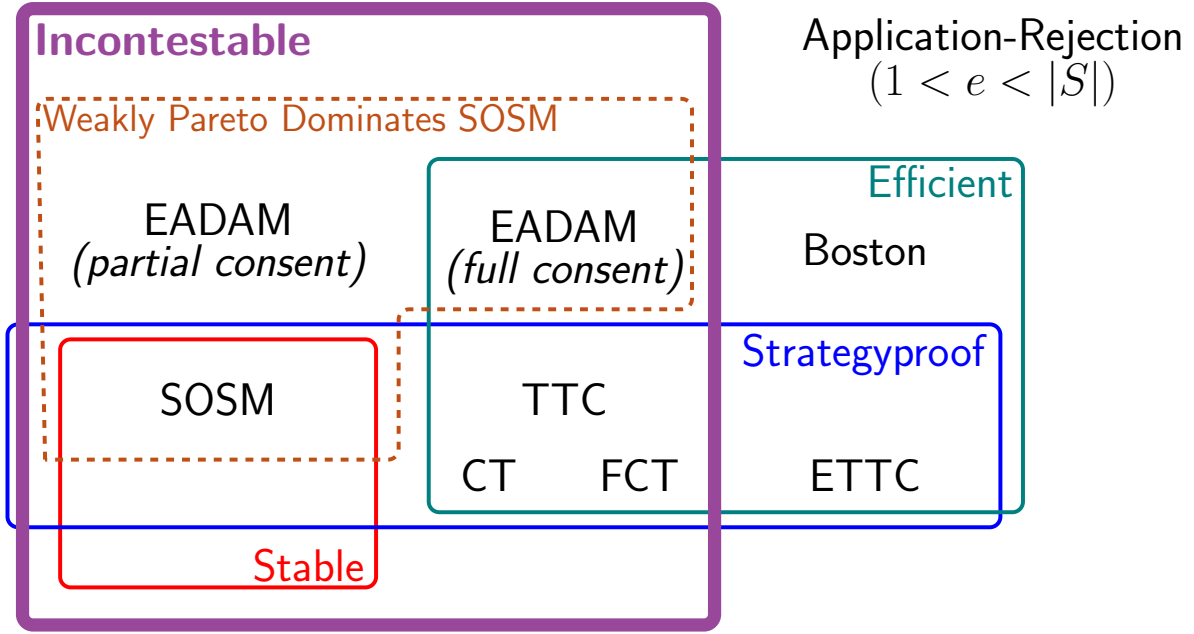

\newpage

\bibliographystyle{apalike}
\bibliography{incontestable}

\begin{thebibliography}{}

\bibitem[Aizerman and Malishevski, 1981]{aizerman1981general}
Aizerman, M. and Malishevski, A. (1981).
\newblock General theory of best variants choice: Some aspects.
\newblock {\em IEEE Transactions on Automatic Control}, 26(5):1030--1040.

\bibitem[Alva and Manjunath, 2019]{alva2019stable}
Alva, S. and Manjunath, V. (2019).
\newblock Stable-dominating rules.
\newblock Technical report, Working paper, University of Ottawa.

\bibitem[Chakraborty et~al., 2010]{chakraborty2010two}
Chakraborty, A., Citanna, A., and Ostrovsky, M. (2010).
\newblock Two-sided matching with interdependent values.
\newblock {\em Journal of Economic Theory}, 145(1):85--105.

\bibitem[Chen and Kesten, 2017]{chen2017chinese}
Chen, Y. and Kesten, O. (2017).
\newblock Chinese college admissions and school choice reforms: A theoretical
  analysis.
\newblock {\em Journal of Political Economy}, 125(1):99--139.

\bibitem[Chen and M{\"o}ller, 2024]{chen2024regret}
Chen, Y. and M{\"o}ller, M. (2024).
\newblock Regret-free truth-telling in school choice with consent.
\newblock {\em Theoretical Economics}, 19(2):635--666.

\bibitem[Decerf and Van~der Linden, 2021]{decerf2021manipulability}
Decerf, B. and Van~der Linden, M. (2021).
\newblock Manipulability in school choice.
\newblock {\em Journal of Economic Theory}, 197:105313.

\bibitem[Dur et~al., 2019]{dur2019school}
Dur, U., Gitmez, A.~A., and Y{\i}lmaz, {\"O}. (2019).
\newblock School choice under partial fairness.
\newblock {\em Theoretical Economics}, 14(4):1309--1346.

\bibitem[Ehlers and Mass{\'o}, 2015]{ehlers2015matching}
Ehlers, L. and Mass{\'o}, J. (2015).
\newblock Matching markets under (in) complete information.
\newblock {\em Journal of Economic Theory}, 157:295--314.

\bibitem[Ehlers and Morrill, 2020]{ehlers2020legal}
Ehlers, L. and Morrill, T. (2020).
\newblock (il) legal assignments in school choice.
\newblock {\em The Review of Economic Studies}, 87(4):1837--1875.

\bibitem[Ergin, 2002]{ergin2002efficient}
Ergin, H.~I. (2002).
\newblock Efficient resource allocation on the basis of priorities.
\newblock {\em Econometrica}, 70(6):2489--2497.

\bibitem[Fernandez, 2018]{fernandez2018deferred}
Fernandez, M.~A. (2018).
\newblock Deferred acceptance and regret-free truth-telling.
\newblock mimeo.

\bibitem[Grigoryan and M{\"o}ller, 2023]{grigoryan2023theory}
Grigoryan, A. and M{\"o}ller, M. (2023).
\newblock A theory of auditability for allocation mechanisms.
\newblock {\em arXiv preprint arXiv:2305.09314}.

\bibitem[Grigoryan and M{\"o}ller, 2024]{grigoryan2024auditability}
Grigoryan, A. and M{\"o}ller, M. (2024).
\newblock Auditability in school choice.
\newblock In {\em AEA Papers and Proceedings}, volume 114, pages 492--496.
  American Economic Association 2014 Broadway, Suite 305, Nashville, TN 37203.

\bibitem[Hakimov and Kesten, 2018]{hakimov2018equitable}
Hakimov, R. and Kesten, O. (2018).
\newblock The equitable top trading cycles mechanism for school choice.
\newblock {\em International Economic Review}, 59(4):2219--2258.

\bibitem[Kagel and Roth, 2000]{kagel2000dynamics}
Kagel, J.~H. and Roth, A.~E. (2000).
\newblock The dynamics of reorganization in matching markets: A laboratory
  experiment motivated by a natural experiment.
\newblock {\em The Quarterly Journal of Economics}, 115(1):201--235.

\bibitem[Kesten, 2010]{kesten2010school}
Kesten, O. (2010).
\newblock School choice with consent.
\newblock {\em The Quarterly Journal of Economics}, 125(3):1297--1348.

\bibitem[Li, 2017]{li2017obviously}
Li, S. (2017).
\newblock Obviously strategy-proof mechanisms.
\newblock {\em American Economic Review}, 107(11):3257--3287.

\bibitem[Liu et~al., 2014]{liu2014stable}
Liu, Q., Mailath, G.~J., Postlewaite, A., and Samuelson, L. (2014).
\newblock Stable matching with incomplete information.
\newblock {\em Econometrica}, 82(2):541--587.

\bibitem[McKinney et~al., 2005]{mckinney2005collapse}
McKinney, C.~N., Niederle, M., and Roth, A.~E. (2005).
\newblock The collapse of a medical labor clearinghouse (and why such failures
  are rare).
\newblock {\em American Economic Review}, 95(3):878--889.

\bibitem[M{\"o}ller, 2022]{moller2022transparent}
M{\"o}ller, M. (2022).
\newblock Transparent matching mechanisms.
\newblock Available at SSRN 4073464.

\bibitem[Morrill, 2015]{morrill2015two}
Morrill, T. (2015).
\newblock Two simple variations of top trading cycles.
\newblock {\em Economic Theory}, 60(1):123--140.

\bibitem[Moulin, 1985]{moulin1985choice}
Moulin, H. (1985).
\newblock Choice functions over a finite set: a summary.
\newblock {\em Social Choice and Welfare}, 2(2):147--160.

\bibitem[Reny, 2022]{reny2022efficient}
Reny, P.~J. (2022).
\newblock Efficient matching in the school choice problem.
\newblock {\em American Economic Review}, 112(6):2025--43.

\bibitem[Roth, 1991]{roth1991natural}
Roth, A.~E. (1991).
\newblock A natural experiment in the organization of entry-level labor
  markets: Regional markets for new physicians and surgeons in the united
  kingdom.
\newblock {\em The American Economic Review}, 81(3):415--440.

\bibitem[Troyan and Morrill, 2020]{troyan2020obvious}
Troyan, P. and Morrill, T. (2020).
\newblock Obvious manipulations.
\newblock {\em Journal of Economic Theory}, 185:104970.

\bibitem[Yenmez, 2013]{yenmez2013incentive}
Yenmez, M.~B. (2013).
\newblock Incentive-compatible matching mechanisms: consistency with various
  stability notions.
\newblock {\em American Economic Journal: Microeconomics}, 5(4):120--141.

\end{thebibliography}


\begin{thebibliography}{}

\bibitem[Chen and Kesten, 2017]{chen2017chinese}
Chen, Y. and Kesten, O. (2017).
\newblock Chinese college admissions and school choice reforms: A theoretical
  analysis.
\newblock {\em Journal of Political Economy}, 125(1):99--139.

\bibitem[Ehlers and Morrill, 2020]{ehlers2020legal}
Ehlers, L. and Morrill, T. (2020).
\newblock (il) legal assignments in school choice.
\newblock {\em The Review of Economic Studies}, 87(4):1837--1875.

\bibitem[Hakimov and Kesten, 2018]{hakimov2018equitable}
Hakimov, R. and Kesten, O. (2018).
\newblock The equitable top trading cycles mechanism for school choice.
\newblock {\em International Economic Review}, 59(4):2219--2258.

\bibitem[Morrill, 2015]{morrill2015two}
Morrill, T. (2015).
\newblock Two simple variations of top trading cycles.
\newblock {\em Economic Theory}, 60(1):123--140.

\bibitem[Reny, 2022]{reny2022efficient}
Reny, P.~J. (2022).
\newblock Efficient matching in the school choice problem.
\newblock {\em American Economic Review}, 112(6):2025--43.

\bibitem[Tang and Yu, 2014]{tang2014new}
Tang, Q. and Yu, J. (2014).
\newblock A new perspective on {K}esten's school choice with consent idea.
\newblock {\em Journal of Economic Theory}, 154:543--561.

\bibitem[Troyan and Morrill, 2020]{troyan2020obvious}
Troyan, P. and Morrill, T. (2020).
\newblock Obvious manipulations.
\newblock {\em Journal of Economic Theory}, 185:104970.

\end{thebibliography}

\newpage

\appendix

\section{Proof of Theorem \ref{thm:new_theorem_1}} 
\label{sec:prooftheorem1}

Suppose that $\mu(i)\notin \Safe^E(\succ_i,\Hi)$. Let
$\hS=\Safe^E(\succ_i,\Hi)$. Then $\underline{s}^E(\succ_i,\Hi)\succ_i \mu(i)$,
and thus $s\succ_i \mu(i)$ for every school $s\in \hS$.
Fix any $(\hC,\hmu)$ compatible with $\Hi$ (so $\hmu(i)=\mu(i)$).
Since $\hS$ is an $\Hi$-high-priority set, there exists $s\in \hS$
such that $i\in \hC_s(\hmu(s)\cup\{i\})$.
Since $s\succ_i \mu(i)$ and $\hmu(i)=\mu(i)$, student $i$ has a
legitimate complaint at $\mu$ given $\Hi$.

\medskip

Suppose that $i$ has a legitimate complaint at $\mu$ given $\Hi$.
Let $T=\{s\in S:\ s\succ_i \mu(i)\}$.
Then $T\neq\varnothing$; otherwise no school would be strictly
preferred to $\mu(i)$, contradicting the existence of a legitimate
complaint. Since $i$ has a legitimate complaint, for every
compatible pair $(\hC,\widehat\mu)$, there exists a school $s\in T$
such that $i\in \hC_s(\widehat\mu(s)\cup\{i\})$. 
Hence $T$ is an $\Hi$-high-priority set for $i$.
Since $\Safe^E(\succ_i,\Hi)$ maximizes the least-preferred school among
all $\Hi$-high-priority sets, we have
$\underline s^E(\succ_i,\Hi)\succeq_i\min_{\succ_i}T$.
But $\min_{\succ_i}T\succ_i\mu(i)$ by construction of $T$.
Therefore $\underline s^E(\succ_i,\Hi)\succ_i\mu(i)$.
Hence $\mu(i)\notin
\{s\in S:\ s\succeq_i\underline s^E(\succ_i,\Hi)\}$,
which means that $\mu(i)\notin\Safe^E(\succ_i,\Hi)$.

\section{Proof of Theorem \ref{thm:equiv-interim-expost-top-set}} 
\label{sec:top_set_then_incontestable}

Fix $\pi\in\Pi$, a problem $(\succ,C)$, an assignment
$\mu\in\mathcal{M}(C)$, and a
student $i$. Let
$\mathcal{K}_i=\pi_i^I(\succ_i,C)$, 
and
$\mathcal H_i=\pi_i^E(\succ_i,C,\mu)
=\{(C',\mu')\in\mathcal K_i\times\mathcal M_i(\mu):
\mu'\in\mathcal M(C')\}$.

\textit{(If)} Suppose that
$\mu(i)\in\Safe^I(\succ_i,\mathcal{K}_i)$. Assume to the contrary that
$\mu(i)\notin\Safe^E(\succ_i,\mathcal{H}_i)$, and let
$T=\Safe^E(\succ_i,\mathcal{H}_i)$. By definition, $T$ is an ex-post
$\mathcal{H}_i$-high-priority set. 

We start by showing that $T$ is also an interim
$\mathcal{K}_i$-high-priority set.

Fix any $C'\in\mathcal K_i$ and any $\mu'\in\mathcal M(C')$.
All capacities in the following construction are those determined by
$C'$. If $\mu'(i)\in T$, then, for $s=\mu'(i)$, since
$|\mu'(s)|\leq q_s(C'_s)$ and all students are acceptable for all
schools, we have $i\in C'_s(\mu'(s)\cup\{i\})$. Suppose instead
that $\mu'(i)\notin T$. Since also $\mu(i)\notin T$, we can
construct an assignment
$\widetilde\mu\in\mathcal M(C')\cap\mathcal M_i(\mu)$ such
that $\widetilde\mu(s)=\mu'(s)$ for every $s\in T$. Indeed, if
$\mu'(i)=\mu(i)$, take $\widetilde\mu=\mu'$. Otherwise, move $i$
from $\mu'(i)$ to $\mu(i)$ and, if $\mu(i)$ is full under $\mu'$
at capacity $q_{\mu(i)}(C'_{\mu(i)})$, move one of its assignees
to the seat vacated by $i$ at $\mu'(i)$. Such an assignee exists
because the capacity is positive. These adjustments preserve
feasibility under $C'$ and concern only schools outside $T$.

Thus, $(C',\widetilde{\mu})\in\mathcal{H}_i$. Since $T$ is an
ex-post $\mathcal{H}_i$-high-priority set, there exists $s\in T$ such
that
\begin{equation*}
i\in C'_s(\widetilde{\mu}(s)\cup\{i\})
=C'_s(\mu'(s)\cup\{i\}).
\end{equation*}
Hence, $T$ is an interim $\mathcal{K}_i$-high-priority set. 

Since $T$
is an upper contour set, the definition of the interim safety-set implies
\begin{equation*}
\Safe^I(\succ_i,\mathcal{K}_i)
\subseteq T
=\Safe^E(\succ_i,\mathcal{H}_i).
\end{equation*}
This contradicts
$\mu(i)\in\Safe^I(\succ_i,\mathcal{K}_i)$ and $\mu(i)\notin T$.

\medskip
\textit{(Only if)} Since
$\mathcal H_i\subseteq
\{(C',\mu'):C'\in\mathcal K_i,\ \mu'\in\mathcal M(C')\}$, 
  every interim
$\mathcal{K}_i$-high-priority set is also an ex-post
$\mathcal{H}_i$-high-priority set. Therefore,
\begin{equation*}
\Safe^E(\succ_i,\mathcal{H}_i)
\subseteq
\Safe^I(\succ_i,\mathcal{K}_i).
\end{equation*}
Thus, $\mu(i)\in\Safe^E(\succ_i,\mathcal{H}_i)$ implies that
$\mu(i)\in\Safe^I(\succ_i,\mathcal{K}_i)$.

Finally, by Theorem~\ref{thm:new_theorem_1} and the definition of
$\Pi$-incontestability, a mechanism is $\Pi$-incontestable if and only
if its assignment belongs to the corresponding ex-post safety-set for
every information rule, problem, and student. Applying the equivalence
above to $\mu=\varphi(\succ,C)$ gives the stated interim safety-set
condition.

\section{Proof of Theorem \ref{thm:attainability}}
\label{sec:prooftheorem2}

The proof of Theorem~\ref{thm:attainability} relies on the following lemmas.

\begin{lemma}[\citeauthor{aizerman1981general},
  \citeyear{aizerman1981general}]\label{lem:revealed_order_PI} 
  Let $C_s:2^I\to 2^I$ be a choice function satisfying path independence.
  Let $\mathcal R_s=\{C_s(X)\,:\,X\subseteq I\}$ be the range of $C_s$.
  Define a binary relation $\unrhd_s$ on $\mathcal R_s$ by
  \begin{equation}
    \label{eq:revealed_pref_range}
    J\unrhd_s J' \quad \Leftrightarrow \quad C_s(J\cup J')=J.
  \end{equation}
  Then $\unrhd_s$ is a partial order on $\mathcal R_s$ (i.e., it is
  reflexive, antisymmetric, and transitive). 
\end{lemma}

\begin{lemma}\label{lemma8}
Let $i$ be a student with interim information set $\Ki$ and preference
list $\succ_i$. Let $C\in\Ki$ be a choice profile such that $C_s$ is
path independent for every $s\in S$. Suppose that, for a non-empty
set of schools $\hS=\{s_1,\dots,s_k\}\subseteq S\setminus\{s_0\}$,
there are pairwise disjoint sets
$J_1,\dots,J_k\subseteq I\setminus\{i\}$ such that, for every
$h=1,\dots,k$,
\begin{equation}\label{eq:Jh_full_reject}
|J_h|=q_{s_h}(C_{s_h})
\qquad\text{and}\qquad
C_{s_h}(J_h\cup\{i\})=J_h.
\end{equation}
Let $\varphi$ be a mechanism satisfying group protection. Then there
exists $\succ_{-i}^\ast$ such that
$\varphi_i((\succ_i,\succ_{-i}^\ast),C)\notin\hS$, and such that,
for every student
$j\notin\varphi_{\hS}((\succ_i,\succ_{-i}^\ast),C)\cup\{i\}$
and every school $s\notin\hS\cup\{s_0\}$, $s_0\succ_j^\ast s$.
\end{lemma}

\begin{proof}
Fix $i$, $\mathcal{K}_i$, a choice profile $C$ compatible with $\Ki$,
a non-empty set
$\hS=\{s_1,\dots,s_k\}\subseteq S\backslash\{s_0\}$, and pairwise
disjoint sets $J_1,\dots,J_k\subseteq I\setminus\{i\}$ satisfying
Eq.~\eqref{eq:Jh_full_reject}.

We first construct an assignment $\mu\in\mathcal M(C)$ such that
$\mu(i)=s_0$ and
\begin{equation}\label{eq:muLemma8_correct}
\mu(s)=
\begin{cases}
C_s(\mu(s)\cup\mu(s_0)) & \text{if }s\in\hS,\\
\varnothing & \text{if }s\notin\hS\cup\{s_0\},\\
I\backslash\bigcup_{h=1}^k\mu(s_h) & \text{if }s=s_0.
\end{cases}
\end{equation}

Let $\mu^0$ be the assignment such that $\mu^0(s_h)=J_h$ for each
$h=1,\dots,k$ and $\mu^0(s)=\varnothing$ for
$s\notin\hS\cup\{s_0\}$, and
$\mu^0(s_0)=I\backslash\cup_h J_h$.
By Eq.~\eqref{eq:Jh_full_reject}, $\mu^0\in\mathcal M(C)$.
Let $Z^0=\mu^0(s_0)$.
Since $i\notin J_h$ for all $h$, we
have $\mu^0(i)=s_0$, hence $i\in Z^0$.
We consider the following multi-step procedure.

\medskip
\noindent\textbf{Step } $\mathbf{t\geq 1}$\quad
Let $\mathcal S^{t}=\{s\in\hS\,:\,
C_s\big(\mu^{t-1}(s)\cup Z^{t-1}\big)\neq\mu^{t-1}(s)\}$.
If $\mathcal S^{t}=\varnothing$, stop and set $\mu=\mu^{t-1}$
and $Z=Z^{t-1}$.
Otherwise, pick any $s^{t}\in\mathcal S^{t}$ and let $\mu^{t}$ be
defined as
\begin{equation*}
\mu^{t}(s)=
\begin{cases}
C_s\big(\mu^{t-1}(s^{t})\cup Z^{t-1}\big)
    & \text{if }s=s^{t},\\
\mu^{t-1}(s) & \text{if }s\neq s^{t},s_0,
\end{cases}
\end{equation*}
and $\mu^t(s_0)=I\backslash\cup_{s\neq s_0}\mu^t(s)$.
Let $Z^t=\mu^t(s_0)$.

\medskip

At each update, the school chooses from its previous assignees and
$Z^{t-1}$, so the schools' sets of assignees remain pairwise disjoint.
By the definition of $q_s(C_s)$, the updated set assigned to each
school $s$ has size at most $q_s(C_s)$. Thus
$\mu^t\in\mathcal M(C)$ at every step.

We first show that the procedure must eventually stop after a finite
number of steps. To that end, consider any step $t$ at which an update
is made, and let $s=s^{t}$,
$B=\mu^{t-1}(s)$ and $A=\mu^{t}(s)=C_s(B\cup Z^{t-1})$.
Since $A\subseteq A\cup B\subseteq B\cup Z^{t-1}$,
consistency yields
$C_s(A\cup B)=C_s(B\cup Z^{t-1})=A$.
Therefore, $A\unrhd_s B$ (where
$\unrhd_s$ is defined in Lemma~\ref{lem:revealed_order_PI}).
Since $A\neq B$, we have $A\rhd_s B$, where $\rhd_s$ denotes the
strict part of $\unrhd_s$.
Since $\mathcal R_s$ is finite for each $s$ and $\rhd_s$ is acyclic,
each school can update its assignment only finitely many times.
Hence the procedure terminates.

Denote by $T$ the last step at which a school updates its
assignment, setting $T=0$ if there is no update.
That is, $\mathcal S^{T+1}=\varnothing$. So, for every
$s\in\hS$, $\mu(s)=\mu^{T}(s)$ and by construction,
$C_s(\mu(s)\cup Z)=\mu(s)$.
Also by construction, $\mu(s)=\varnothing$ for each
$s\notin\hS\cup\{s_0\}$, and $\mu(s_0)=I\backslash\mu(\hS)$.
This is exactly Eq.~\eqref{eq:muLemma8_correct}.

For notational convenience, set $\mu^t=\mu^T$, $Z^t=Z^T$, and
$\mathcal S^t=\varnothing$ for all $t>T$.

It remains to show that $i\in Z^{T}$.
We prove a stronger claim by induction on $t$: for every $t\geq 0$,
\begin{equation}\label{eq:i_in_Zt_invariant}
i\in Z^{t}
\qquad\text{and}\qquad
i\notin C_s\big(\mu^{t}(s)\cup\{i\}\big)
\quad\text{for all }s\in\hS.
\end{equation}
If this claim holds, then since $Z^{t}=\mu^{t}(s_0)$, the first part
of Eq.~\eqref{eq:i_in_Zt_invariant} implies
that $i$ is never assigned to a school in $\hS$ at any step.
In particular, $i\in Z^{T}$, as desired.

We first show that Eq.~\eqref{eq:i_in_Zt_invariant} holds for $t=0$.
By construction, $\mu^0(s_h)=J_h$ for each $h=1,\dots,k$, and
$i\notin J_h$ since $J_h\subseteq I\setminus\{i\}$.
Hence $i\notin\mu^0(s)$ for every
$s\neq s_0$, so $i\in Z^0=\mu^0(s_0)$.
Moreover, for each $s_h\in\hS$,
Eq.~\eqref{eq:Jh_full_reject} gives
$C_{s_h}\big(\mu^0(s_h)\cup\{i\}\big)
=C_{s_h}(J_h\cup\{i\})=J_h=\mu^0(s_h)$.
So, $i\notin C_{s_h}(\mu^0(s_h)\cup\{i\})$.
For schools $s\notin\hS\cup\{s_0\}$,
$\mu^0(s)=\varnothing$ and they never update.
Thus Eq.~\eqref{eq:i_in_Zt_invariant} holds at $t=0$.

\medskip
Fix $t\geq 1$ and assume Eq.~\eqref{eq:i_in_Zt_invariant}
holds at $t-1$.
If $\mathcal S^{t}=\varnothing$, then $\mu^{t}=\mu^{t-1}$ and
$Z^{t}=Z^{t-1}$, so Eq.~\eqref{eq:i_in_Zt_invariant}
also holds at $t$.
Suppose instead that $\mathcal S^{t}\neq\varnothing$ and the procedure
updates school $s^{t}\in\mathcal S^{t}$.
Let $M\equiv\mu^{t-1}(s^{t})\cup Z^{t-1}$.
By the induction hypothesis, $i\in Z^{t-1}$, hence
$\mu^{t-1}(s^{t})\cup\{i\}\subseteq M$.
Also by the induction hypothesis,
$i\notin C_{s^{t}}(\mu^{t-1}(s^{t})\cup\{i\})$, i.e.,
$i\in R_{s^{t}}(\mu^{t-1}(s^{t})\cup\{i\})$.
By substitutability,
$R_{s^{t}}(\mu^{t-1}(s^{t})\cup\{i\})\subseteq R_{s^{t}}(M)$,
so $i\in R_{s^{t}}(M)$ and therefore
$i\notin C_{s^{t}}(M)=\mu^{t}(s^{t})$.
No other school $s\neq s^{t},s_0$ changes its assigned set at step $t$,
so $i\notin\mu^{t}(s)$ for every $s\neq s_0$.
Hence, by definition of
$Z^{t}=\mu^{t}(s_0)=I\setminus\bigcup_{s\neq s_0}\mu^{t}(s)$,
we obtain $i\in Z^{t}$.

It remains to verify the second part of
Eq.~\eqref{eq:i_in_Zt_invariant} at step $t$.
For any school $s\in\hS$ with $s\neq s^{t}$ we have
$\mu^{t}(s)=\mu^{t-1}(s)$, so
$i\notin C_s(\mu^{t}(s)\cup\{i\})$ follows
from the induction hypothesis.
For $s=s^{t}$, note that
$\mu^{t}(s^{t})=C_{s^{t}}(M)\subseteq M$ and $i\in M$, so
$\mu^{t}(s^{t})\subseteq\mu^{t}(s^{t})\cup\{i\}\subseteq M$.
By consistency, since $C_{s^{t}}(M)=\mu^{t}(s^{t})$ and
$\mu^{t}(s^{t})\subseteq\mu^{t}(s^{t})\cup\{i\}\subseteq M$,
we have
$C_{s^{t}}\big(\mu^{t}(s^{t})\cup\{i\}\big)
=C_{s^{t}}(M)=\mu^{t}(s^{t})$,
and thus $i\notin C_{s^{t}}(\mu^{t}(s^{t})\cup\{i\})$.
This completes the induction step and proves
Eq.~\eqref{eq:i_in_Zt_invariant} for all $t$.
Taking $t=T$ yields $i\in Z^{T}$, as claimed.

We now use this assignment to apply group protection.
For each $s\in\hS$, let $G_s=\mu(s)$, and let
\begin{equation*}
G=\bigcup_{s\in\hS}G_s,
\qquad Z=\mu(s_0)=I\setminus G.
\end{equation*}
By Eq.~\eqref{eq:muLemma8_correct}, $C_s(G_s\cup Z)=G_s$
for every $s\in\hS$.
Moreover, Eq.~\eqref{eq:i_in_Zt_invariant} at $t=T$ implies
that $i\in Z$ and
$i\notin C_s(G_s\cup\{i\})$ for every $s\in\hS$.
If $|G_s|<q_s(C_s)$, the assumption that no school deems any
student unacceptable would imply
$C_s(G_s\cup\{i\})=G_s\cup\{i\}$, a contradiction.
Since $\mu\in\mathcal M(C)$, we therefore have
$|G_s|=q_s(C_s)$ for every $s\in\hS$.

For every $s\in\hS$ and every $j\in G_s$, let $\succ_j^\ast$
rank $s$ first, followed by the other schools in $\hS$ in any
order, then $s_0$, and then the remaining schools in any order.
For every $j\in Z\setminus\{i\}$, let $\succ_j^\ast$ rank $s_0$
first and all other schools in any order.
Student $i$'s preference list remains $\succ_i$.

At the problem $((\succ_i,\succ_{-i}^\ast),C)$, the sets
$(G_s)_{s\in\hS}$ satisfy all the conditions of group protection.
Consequently,
\begin{equation*}
\varphi_s((\succ_i,\succ_{-i}^\ast),C)\subseteq G
\qquad\text{for every }s\in\hS.
\end{equation*}
Since $i\notin G$, it follows that
$\varphi_i((\succ_i,\succ_{-i}^\ast),C)\notin\hS$.
Finally, every student $j\neq i$ ranks $s_0$ above every school
outside $\hS\cup\{s_0\}$.
This implies the last claim of the lemma.
\end{proof}

\begin{proofof}{thm:attainability}
Since $\varphi$ is $\Pi^C$-incontestable,
Corollary~\ref{cor:Pi-C_incontestable_iff_respect_top_sets} implies
that it respects safety-sets. Fix a student $i$ with preference
list $\succ_i$ and interim information set $\Ki$, and let
$\hS=\Safe^I(\succ_i,\Ki)$.
For every $C'\in\Ki$ and every $\succ_{-i}'$, respect for
safety-sets and their monotonicity imply
\begin{equation*}
\varphi_i((\succ_i,\succ_{-i}'),C')
\in\Safe^I(\succ_i,\{C'\})
\subseteq\Safe^I(\succ_i,\Ki)=\hS.
\end{equation*}
Thus every attainable school belongs to $\hS$.

Recall also that $C_{s_0}(J)=J$ for every $J\subseteq I$.
Hence the upper contour set of $s_0$ is an interim high-priority
set for every student and every interim information set. Every
interim safety-set is therefore contained in the corresponding
upper contour set of $s_0$. In particular, $\varphi$ is individually
rational at every problem, and
$\hS\subseteq\{s\in S:s\succeq_i s_0\}$.

It remains to show that every $s^\ast\in\hS$ is attainable.
Fix such an $s^\ast$. We will construct a choice profile $C\in\Ki$
and a preference profile $\succ_{-i}^\ast$ such that
$\varphi_i((\succ_i,\succ_{-i}^\ast),C)=s^\ast$.

\medskip
\noindent\textit{Case 1: $s^\ast=s_0$.}\quad
Let $\tS=\{s\in S:s\succ_i s_0\}$.
If $\tS=\varnothing$, individual rationality implies that $i$ is
assigned to $s_0$ at every compatible problem, as desired.
Suppose that $\tS\neq\varnothing$. Since $s_0\in\hS$ and $\hS$
is the smallest upper contour set that is an interim
$\Ki$-high-priority set, $\tS$ is not an interim
$\Ki$-high-priority set. Therefore, there exist $C\in\Ki$ and
$\nu\in\mathcal M(C)$ such that
\begin{equation*}
i\notin C_s(\nu(s)\cup\{i\})
\qquad\text{for every }s\in\tS.
\end{equation*}
Fix $s\in\tS$. Feasibility and the assumption that no school
deems any student unacceptable imply $C_s(\nu(s))=\nu(s)$.
Thus $i\notin\nu(s)$, since otherwise $i$ would be chosen from
$\nu(s)\cup\{i\}=\nu(s)$.
The same assumption implies that $|\nu(s)|=q_s(C_s)$: if
$|\nu(s)|<q_s(C_s)$, school $s$ would choose $i$ from
$\nu(s)\cup\{i\}$. Finally,
$C_s(\nu(s)\cup\{i\})\subseteq\nu(s)$, so consistency gives
\begin{equation*}
C_s(\nu(s)\cup\{i\})=C_s(\nu(s))=\nu(s).
\end{equation*}
The sets $(\nu(s))_{s\in\tS}$ are pairwise disjoint and exclude
$i$. Moreover, every choice function in $C$ is path independent
by the domain assumption of the theorem. The hypotheses of
Lemma~\ref{lemma8} are therefore satisfied with $\tS$.
It follows that there exists $\succ_{-i}^\ast$ such that
$\varphi_i((\succ_i,\succ_{-i}^\ast),C)\notin\tS$.
Individual rationality then implies
$\varphi_i((\succ_i,\succ_{-i}^\ast),C)=s_0=s^\ast$.

\medskip
\noindent\textit{Case 2: $s^\ast\neq s_0$ and $s^\ast$ is $i$'s most
preferred school.}\quad
Choose any $C\in\Ki$. Let $\succ_{-i}^\ast$ be such that
$s_0\succ_j^\ast s^\ast$ for every $j\neq i$.
Individual rationality implies that no student other than $i$
can be assigned to $s^\ast$. If $i$ were not assigned to $s^\ast$,
that school would be empty, while $i$ would strictly prefer it to
her assignment. Since $q_{s^\ast}(C_{s^\ast})>0$, this would
contradict non-wastefulness. Hence
$\varphi_i((\succ_i,\succ_{-i}^\ast),C)=s^\ast$.

\medskip
\noindent\textit{Case 3: $s^\ast\neq s_0$ and $s^\ast$ is not $i$'s
most preferred school.}\quad
Let $\tS=\{s\in S:s\succ_i s^\ast\}$.
Then $\tS\neq\varnothing$. Since
$s^\ast\in\hS\subseteq\{s\in S:s\succeq_i s_0\}$ and
$s^\ast\neq s_0$, we have $s^\ast\succ_i s_0$, so
$s_0\notin\tS$. Moreover, $\tS$ is a proper subset of $\hS$
and an upper contour set. Thus it is not an interim
$\Ki$-high-priority set.
There consequently exist $C\in\Ki$ and $\nu\in\mathcal M(C)$
such that $i\notin C_s(\nu(s)\cup\{i\})$ for every $s\in\tS$.
As in Case 1, the sets $(\nu(s))_{s\in\tS}$ are pairwise disjoint,
exclude $i$, and satisfy
\begin{equation*}
|\nu(s)|=q_s(C_s)
\qquad\text{and}\qquad
C_s(\nu(s)\cup\{i\})=\nu(s)
\qquad\text{for every }s\in\tS.
\end{equation*}
By Lemma~\ref{lemma8}, there exists $\succ_{-i}^\ast$ such that,
writing $\mu=\varphi((\succ_i,\succ_{-i}^\ast),C)\in\mathcal M(C)$,
we have $\mu(i)\notin\tS$ and
$s_0\succ_j^\ast s$ for every
$j\notin\mu(\tS)\cup\{i\}$ and every
$s\notin\tS\cup\{s_0\}$.

We now show that no student other than $i$ can be assigned to
$s^\ast$. Suppose, toward a contradiction, that some $j\neq i$
is assigned to $s^\ast$. Since $s^\ast\notin\tS$, this student
is not in $\mu(\tS)$. The preference condition supplied by the
lemma therefore implies $s_0\succ_j^\ast s^\ast$, contradicting
individual rationality.
If $i$ were not assigned to $s^\ast$, this school would consequently
be empty. Since $\mu(i)\notin\tS$ and $\mu(i)\neq s^\ast$,
we would have $s^\ast\succ_i\mu(i)$. Together with
$q_{s^\ast}(C_{s^\ast})>0$, this contradicts non-wastefulness.
Thus $\mu(i)=s^\ast$.

In each case, $s^\ast$ is assigned to $i$ at a problem compatible
with $\Ki$. This proves that every school in $\hS$ is attainable.
\end{proofof}

\section{Proof of Proposition~\ref{prop:family-incontestable-mech}}
\label{sec:proof_prop_family_incontestable}

\subsection{Any mechanism that Pareto-dominates a
  $\Pi^C$-incontestable mechanism}
\label{sec:Pareto_dominates_then_incontestable}

For that result, we use the following generalization of the Rural
Hospital Theorem (See \cite{alva2019stable} for a related result). 

\begin{theorem}
\label{thm:generalized-HRT}
  Fix a school choice problem $(\succ,C)$ with a finite set of
  students $I$ and schools $S$.
  Each school $s\in S$ has capacity $q_s=q_s(C_s)\in\mathbb{N}$
  and a choice function $C_s:2^I\to 2^I$ satisfying the following
  acceptability condition: for every $J\subseteq I$ with
  $|J|<q_s$ and every $i\in I\setminus J$,
  $i\in C_s(J\cup\{i\})$ (i.e., a school cannot reject a student
  when it has a vacant seat).
  Let $\mu\in\mathcal M(C)$ be an assignment that Pareto dominates
  an individually rational and non-wasteful assignment
  $\mu'\in\mathcal M(C)$. Then the following holds:
  \begin{itemize}
  \item[\textit{(i)}] for each school $s\in S$, $|\mu(s)|=|\mu'(s)|$.
    
  \item[\textit{(ii)}] If  $|\mu'(s)|<q_s$ then $\mu(s)=\mu'(s)$. 

  \item[\textit{(iii)}] for each student $i\in I$, $\mu(i)\in S\backslash\{s_0\}$ if,
    and only if $\mu'(i)\in S\backslash\{s_0\}$. 
  \end{itemize}
\end{theorem}

  Given two assignments $\mu$ and $\mu'$, let $S^\ast=S\setminus\{s_0\}$.
For each real school $s\in S^\ast$, let $E_s^{\mu,\mu'}$ and
$L_s^{\mu,\mu'}$ denote the students who \emph{enter} and \emph{leave}
school $s$ when moving from $\mu'$ to $\mu$, respectively:
\begin{align*}
  E_s^{\mu,\mu'} &:= \mu(s)\setminus \mu'(s),\\
  L_s^{\mu,\mu'} &:= \mu'(s)\setminus \mu(s).
\end{align*}
The proof shows that for each school $s$, $|E_s^{\mu,\mu'}|=|L_s^{\mu,\mu'}|$.

\begin{proof}
Let $\mu$ weakly Pareto dominate $\mu'$. Define
\begin{equation*}
  I^\Delta := \{i\in I:\mu(i)\neq \mu'(i)\}
  \qquad\text{and}\qquad
  \overline{S} := \{s\in S^\ast:|\mu'(s)|=q_s\}.
\end{equation*}

We first show that any real school that gains students must be full
under $\mu'$. We claim that, for each $s\in S^\ast$,
\begin{equation*}
  E_s^{\mu,\mu'}\neq \emptyset \ \Rightarrow\ s\in \overline{S}.
\end{equation*}
Take $s\in S^\ast$ with $E_s^{\mu,\mu'}\neq \emptyset$ and let
$i\in E_s^{\mu,\mu'}$. Then $\mu(i)=s$ and $\mu'(i)\neq s$.
Since $\mu$ weakly Pareto dominates $\mu'$ and preferences are strict,
we have $s\succ_i \mu'(i)$. If $|\mu'(s)|<q_s$, then $s$ has a vacant
seat under $\mu'$ and, by the acceptability condition,
$i\in C_s(\mu'(s)\cup\{i\})$. Thus $\mu'$ is wasteful, contradicting
the assumption that $\mu'$ is non-wasteful. Hence $|\mu'(s)|=q_s$.
Equivalently,
\begin{equation}\label{eq:step1}
  s\notin \overline{S}\ \Rightarrow\ E_s^{\mu,\mu'}=\emptyset.
\end{equation}

We next show that no student can move from a real school to the null
school. Suppose that $\mu(i)=s_0$ and $\mu'(i)\in S^\ast$. Since
$\mu'$ is individually rational, $\mu'(i)\succ_i s_0$. This contradicts
the fact that $\mu$ weakly Pareto dominates $\mu'$. Hence
\begin{equation}\label{eq:no_real_to_null}
  \mu(i)=s_0 \ \Rightarrow\ \mu'(i)=s_0.
\end{equation}
It follows that every student whose assignment changes is assigned to
some real school under $\mu$. Therefore the sets
$\{E_s^{\mu,\mu'}\}_{s\in S^\ast}$ are pairwise disjoint and every
student in $I^\Delta$ belongs to exactly one of them. Hence
\begin{equation*}
  \sum_{s\in S^\ast}|E_s^{\mu,\mu'}|=|I^\Delta|.
\end{equation*}
By \eqref{eq:step1}, only schools in $\overline S$ can have entrants,
so
\begin{equation}\label{eq:entrants}
  \sum_{s\in \overline{S}}|E_s^{\mu,\mu'}|=|I^\Delta|.
\end{equation}

On the other hand, every student who leaves a real school must have a
changed assignment. Since the sets
$\{L_s^{\mu,\mu'}\}_{s\in S^\ast}$ are pairwise disjoint, we have
\begin{equation}\label{eq:leavers_ub}
  \sum_{s\in S^\ast}|L_s^{\mu,\mu'}|\le |I^\Delta|.
\end{equation}
The inequality may be strict only if some student moves from the null
school to a real school.

We now show that full real schools must have at least as many leavers
as entrants. Fix $s\in\overline S$, so $|\mu'(s)|=q_s$. Since
\begin{equation*}
  \mu(s)=\big(\mu'(s)\setminus L_s^{\mu,\mu'}\big)
  \cup E_s^{\mu,\mu'}
  \qquad\text{(disjoint union),}
\end{equation*}
we have
\begin{equation*}
  |\mu(s)|=|\mu'(s)|-|L_s^{\mu,\mu'}|+|E_s^{\mu,\mu'}|.
\end{equation*}
Because $|\mu(s)|\le q_s=|\mu'(s)|$, it follows that
\begin{equation}\label{eq:full_school_ineq}
  |E_s^{\mu,\mu'}|\le |L_s^{\mu,\mu'}|
  \qquad\text{for all }s\in\overline S.
\end{equation}

Combining \eqref{eq:entrants}, \eqref{eq:full_school_ineq}, and
\eqref{eq:leavers_ub}, we obtain
\begin{equation*}
  |I^\Delta|
  =
  \sum_{s\in\overline S}|E_s^{\mu,\mu'}|
  \le
  \sum_{s\in\overline S}|L_s^{\mu,\mu'}|
  \le
  \sum_{s\in S^\ast}|L_s^{\mu,\mu'}|
  \le
  |I^\Delta|.
\end{equation*}
Hence all inequalities are equalities. In particular,
\begin{equation*}
  |E_s^{\mu,\mu'}|=|L_s^{\mu,\mu'}|
  \qquad\text{for every }s\in\overline S.
\end{equation*}
Therefore, for every $s\in\overline S$,
\begin{equation*}
  |\mu(s)|
  =
  |\mu'(s)|-|L_s^{\mu,\mu'}|+|E_s^{\mu,\mu'}|
  =
  |\mu'(s)|
  =
  q_s.
\end{equation*}

Now let $s\in S^\ast\setminus \overline S$. By \eqref{eq:step1},
$E_s^{\mu,\mu'}=\emptyset$. Moreover, since all inequalities above are
equalities, we have
\begin{equation*}
  \sum_{s\in S^\ast\setminus\overline S}|L_s^{\mu,\mu'}|=0.
\end{equation*}
Thus $L_s^{\mu,\mu'}=\emptyset$ for every
$s\in S^\ast\setminus\overline S$. Hence
\begin{equation*}
\mu(s)=\mu'(s)
\qquad\text{for every }s\in S^\ast\setminus\overline S.
\end{equation*}
This proves \textit{(ii)} for all real schools. Together with the
previous paragraph, it also proves \textit{(i)} for all real schools.

It remains to prove that the same students are assigned to real schools
under $\mu$ and $\mu'$. One implication was already shown in
\eqref{eq:no_real_to_null}: if $\mu(i)=s_0$, then $\mu'(i)=s_0$.
For the converse, suppose that $\mu'(i)=s_0$ and $\mu(i)\neq s_0$.
Then $i\in I^\Delta$, but $i$ does not belong to any
$L_s^{\mu,\mu'}$ with $s\in S^\ast$, since she was not assigned to a
real school under $\mu'$. This contradicts the equality
\begin{equation*}
\sum_{s\in S^\ast}|L_s^{\mu,\mu'}|=|I^\Delta|,
\end{equation*}
which implies that every student whose assignment changes must leave
some real school. Therefore,
\begin{equation*}
\mu'(i)=s_0 \ \Rightarrow\ \mu(i)=s_0.
\end{equation*}
Combining the two implications gives
\begin{equation*}
\mu(i)\in S^\ast
\quad\Longleftrightarrow\quad
\mu'(i)\in S^\ast,
\end{equation*}
which proves \textit{(iii)}.

Finally, \textit{(iii)} implies $\mu(s_0)=\mu'(s_0)$. Hence
$|\mu(s_0)|=|\mu'(s_0)|$, so \textit{(i)} also holds for the null
school. Moreover, if $|\mu'(s_0)|<q_{s_0}$, then
$\mu(s_0)=\mu'(s_0)$, so \textit{(ii)} also holds for the null school.
This completes the proof.
\end{proof}

\begin{lemma}
\label{prop:pareto_dom_incontestable_pointwise}
Fix a problem $(\succ,C)$ and let $\mu$ and $\mu'$ be assignments.
If $\mu$ weakly Pareto dominates $\mu'$ and $\mu'$ respects safety-sets at
$(\succ,C)$, then $\mu$ respects safety-sets at $(\succ,C)$.
Moreover, if $\mu'$ is non-wasteful at $(\succ,C)$, then $\mu$ is
non-wasteful at $(\succ,C)$.
\end{lemma}

\begin{proof}
  Fix a problem $(\succ,C)$ and let $q_s=q_s(C_s)$ for each $s\in S$,
and let $\mu,\mu'\in \mathcal{M}(C)$. 
  Let $i\in I$ and let
$\hS_i=\Safe^I(\succ_i,\{C\})$. Since $\mu'$ respects safety-sets,
$\mu'(i)\in \hS_i$. By definition, $\hS_i$ is an upper contour set: if
$s\in\hS_i$ and $t\succeq_i s$, then $t\in\hS_i$. Since $\mu$ weakly
Pareto dominates $\mu'$, we have $\mu(i)\succeq_i\mu'(i)$, and hence
$\mu(i)\in\hS_i$. Since $i$ was arbitrary, $\mu$ respects safety-sets at
$(\succ,C)$.

Assume now that $\mu'$ is non-wasteful, and suppose, toward a
contradiction, that $\mu$ is wasteful. Then there exist a student $i$
and a school $s$ such that
$s\succ_i \mu(i)$ and  $|\mu(s)|<q_s$.
Since $\mu$ weakly Pareto dominates $\mu'$, we have
$\mu(i)\succeq_i \mu'(i)$, and therefore $s\succ_i \mu'(i)$.
Moreover, since $\mu$ weakly Pareto dominates $\mu'$ and $\mu'$ is
non-wasteful, Theorem~\ref{thm:generalized-HRT} implies
$|\mu(s)|=|\mu'(s)|$. Hence $|\mu'(s)|<q_s$, contradicting the
non-wastefulness of $\mu'$.
\end{proof}

Fix a problem $(\succ,C)$ and let
$\mu=\varphi(\succ,C)$ and $\mu'=\psi(\succ,C)$. Since $\psi$ is
$\Pi^C$-incontestable, 
$\psi$  respects safety-sets, and thus $\mu'(i)\in \Safe^I(\succ_i,\{C\})$
for each $i\in I$. By assumption, $\mu$ weakly Pareto
dominates $\mu'$. Lemma
\ref{prop:pareto_dom_incontestable_pointwise} therefore implies that
$\mu$ respects safety-sets at $(\succ,C)$. Since $(\succ,C)$ was
arbitrary, Corollary~\ref{cor:Pi-C_incontestable_iff_respect_top_sets} implies that
$\varphi$ is $\Pi^C$-incontestable.
If $\psi$ is non-wasteful, then $\mu'$ is non-wasteful at every
problem. The non-wastefulness of $\varphi$ follows from the second part
of Lemma~\ref{prop:pareto_dom_incontestable_pointwise}.

\subsection{TTC is $\Pi^C$-incontestable}
\label{sec:ttc-pic-incont}

  For the Top Trading Cycles mechanism, we assume that schools have
responsive priorities. Fix any problem $(\succ,C)$ and a responsive
representation $(r,q(C))$ of its choice profile, with the convention
that $r_s(i)<r_s(j)$ means that $i$ has higher priority than $j$ at $s$.
Throughout the proof, $C$, its priority representation, and its
capacities $q_s=q_s(C_s)$ are held fixed. Since the problem is
arbitrary, capacities may vary across choice profiles in the domain.

\begin{lemma}\label{lemma:TTC_incontestable}
  The Top-Trading Cycles (TTC) mechanism is $\Pi^C$-incontestable.
\end{lemma}

\begin{proof}
By Corollary~\ref{cor:Pi-C_incontestable_iff_respect_top_sets}, it
suffices to show that TTC respects safety-sets.
Let $\mu=\mathrm{TTC}(\succ,C)\in\mathcal M(C)$, fix a student
$i$, and let $\hS=\Safe^I(\succ_i,\{C\})$.
Suppose, toward a contradiction, that $\mu(i)\notin\hS$.
If $s_0\in\hS$, individual rationality of TTC already contradicts
this supposition, since $\hS$ is an upper contour set. Hence
$s_0\notin\hS$.

  By definition of the safety-set, every school in $\hS$ is strictly
  preferred by $i$ to every school 
  in $S\setminus \hS$; hence $s\succ_i \mu(i)$ for all $s\in\hS$.

  Let $k_i$ be the step of the TTC algorithm at which student $i$ is
  removed (and thus assigned). 
  Let $k_{\hS}$ be the step at which the \emph{last remaining seat} among schools in $\hS$ is
  allocated.\footnote{Equivalently: at the beginning of step $k_{\hS}$, at least one school in
    $\hS$ still has an available seat, and at the beginning of step $k_{\hS}+1$ all schools in
    $\hS$ are full.}
  Since student $i$ points at each step to her most preferred school among those that still have
  an available seat, and since every school in $\hS$ is preferred to every school in
  $S\setminus\hS$, student $i$ must point to a school in $\hS$ at every step prior to $k_{\hS}$.
  Therefore, if $\mu(i)\notin\hS$, student $i$ can only be assigned after all schools in $\hS$
  have filled their capacities, that is,
  \begin{equation}\label{eq:TTC_k_i_gt_k_S}
    k_i>k_{\hS}.
  \end{equation}

  For each $s\in\hS$, let $J_s$ be the set of students to whom
$s$ points in the rounds in which one of its seats is allocated.
Every such student belongs to the executed cycle and is removed
in that round. These students are distinct across allocated seats,
both within a round and across rounds. Thus the sets $(J_s)_{s\in\hS}$
are pairwise disjoint and $|J_s|=q_s$ for every $s\in\hS$.
All those seats are allocated before $i$ is removed, by
Eq.~\eqref{eq:TTC_k_i_gt_k_S}. Since a school points to its
highest-priority remaining student, every member of $J_s$ has
higher priority at $s$ than $i$.

Define $\nu$ by $\nu(s)=J_s$ for $s\in\hS$,
$\nu(s)=\varnothing$ for $s\notin\hS\cup\{s_0\}$, and
$\nu(s_0)=I\setminus\bigcup_{s\in\hS}J_s$.
Then $\nu\in\mathcal M(C)$. Responsiveness implies
\begin{equation*}
i\notin C_s(\nu(s)\cup\{i\})
\qquad\text{for every }s\in\hS,
\end{equation*}
contradicting that $\hS=\Safe^I(\succ_i,\{C\})$ is an interim
$\{C\}$-high-priority set. This contradiction shows that
$\mu(i)\in\hS$. Since $i$ was arbitrary, $\mu$ respects safety-sets
at $(\succ,C)$, and thus TTC is $\Pi^C$-incontestable.

\end{proof}

\section{Proof of Proposition~\ref{prop:maxmin-truthful}}
\label{sec:maxmin-incentives}

Fix $i$, $\succ_i$, $\Ki$, $\varphi$, and an alternative preference list
$\succ_i'\in\mathcal P_i$.
Since $\varphi$ is $\Pi^C$-incontestable,
$\varphi$ respects safety-sets (Corollary \ref{cor:Pi-C_incontestable_iff_respect_top_sets}). Let
$\hS=\Safe^I(\succ_i,\Ki)$ and 
$\tS=\Safe^I(\succ_i',\Ki)$.
By Theorem~\ref{thm:attainability},
when $i$ reports $\succ_i$ truthfully, the set of schools that $i$ can
obtain across problems compatible with $\Ki$ is exactly
$\hS$. Hence,
\begin{equation}\label{eq:w_truth}
w_i(\succ_i\,|\,\succ_i,\varphi,\Ki)
=\min_{\succ_i}\widehat{S} =
\underline{s}^I(\succ_i,\Ki),
\end{equation}
\noindent and applying the same theorem when $i$ reports $\succ_i'$,
\begin{equation}\label{eq:w_misreport}
w_i(\succ_i'\mid\succ_i,\varphi,\mathcal K_i)
=\min_{\succ_i}\widetilde S.
\end{equation}
Let $t=\min_{\succ_i}\widetilde S$. The set $\widetilde S$ is
interim $\mathcal K_i$-high-priority, independently of the
preference ordering used to select it. Since
$\widetilde S\subseteq U_i(t)$, monotonicity implies that
$U_i(t)$ is also interim $\mathcal K_i$-high-priority.
The definition of $\widehat S$ as the smallest such upper contour
set gives $\widehat S\subseteq U_i(t)$. Therefore
$\min_{\succ_i}\widehat S\succeq_i t$. Combining
Eqs.~\eqref{eq:w_truth} and~\eqref{eq:w_misreport} yields
\begin{equation*}
w_i(\succ_i\mid\succ_i,\varphi,\mathcal K_i)
\succeq_i
w_i(\succ_i'\mid\succ_i,\varphi,\mathcal K_i),
\end{equation*}
\noindent as required.

\end{document}


\title{Incontestable Assignments\\
Online Appendix
}

\author{Benoit Decerf\footnote{Universit\'e de Namur: benoit.decerf@unamur.be}\and
  Guillaume Haeringer\footnote{Baruch College: Guillaume.Haeringer@baruch.cuny.edu}
\and Martin Van der Linden\footnote{Independent scholar: martinvdl@gmail.com}}

\date{\today
}

\maketitle

\tableofcontents

\newpage

\section{Additional $\Pi^C$-incontestable mechanisms}
\label{sec:addit-pic-incont}

Both Clinch and Trade (CT) and First Clinch and Trade (FCT) are
individually rational and non-wasteful---see \citet{morrill2015two}.

\subsection{Clinch and Trade mechanism}
\label{sec:clinch-trade-mech-1}

Throughout this section, when the choice profile $C$ is induced by
responsive priorities $r$ and capacities $q$, we use the priority
representation $(r,q)$ to describe the algorithms, where
$q_s=q_s(C_s)$. In both algorithms, a real school is available if
it has positive remaining capacity, and the null school $s_0$ is
always available.

The Clinch and Trade mechanism uses the following iterative algorithm.
At each round, given the remaining students and capacities, perform
the two steps below.

\begin{itemize}
\item[\textbf{Step 1}] In the first round, every student is eligible
  to participate in the clinching procedure. In each subsequent
  round, an unassigned student is eligible if the school she
  pointed to in Step 2 of the previous round exhausted its
  capacity in that step.

  Let $q_s^0$ be the current remaining capacity of real school $s$.
  At each clinching subround $h=0,1,\ldots$, each unassigned
  student points to her most preferred available school.
  An eligible student $i$ clinches a real school $s$ if she points
  to $s$ and is among the $q_s^h$ highest-priority unassigned
  students at $s$, that is, if
  \begin{equation*}
  \left|\{j\text{ unassigned}:r_s(j)<r_s(i)\}\right|<q_s^h.
  \end{equation*}
  All clinching students are assigned to the schools they point to
  and removed. Every student pointing to $s_0$ is assigned to
  $s_0$ and removed, whether or not she is eligible to clinch a
  real school. These assignments are determined simultaneously.
  If $m_s^h$ students clinch school $s$ at subround $h$, set
  $q_s^{h+1}=q_s^h-m_s^h$.

  Step 1 ends when no student can be assigned in this step.

\item[\textbf{Step 2}] If students remain, each of them points to
  her most preferred available school, and each available real
  school points to the remaining student with the highest priority.
  There is then at least one cycle. If a student is in a cycle,
  assign her to the school she points to. Remove
  all assigned students and reduce the capacity of each school
  in a cycle by one. All cycles are processed simultaneously.
\end{itemize}

Repeat these two steps until all students are assigned.

\begin{lemma}\label{lemma:CT-incontestable}
The Clinch and Trade mechanism respects safety-sets.
\end{lemma}

\begin{proof}
Fix a problem $(\succ,C)$ and a student $i$. Let
$\widehat S=\Safe^I(\succ_i,\{C\})$.
We show that CT assigns $i$ to a school in $\widehat S$.
Write $q_s=q_s(C_s)$. In the residual problems below, choice
functions use the restricted priorities and select students up to
the remaining numbers of seats. If this number is at least the
number of remaining students, the school chooses all remaining
students and its singleton is a high-priority set.

We repeatedly use the following property of safety-sets: if
$s\in \widehat S$ and $s'\succ_i s$, then $s'\in \widehat S$.
If $s_0\in\widehat S$, individual rationality of CT and this
property give the desired conclusion. Hence suppose that
$s_0\notin\widehat S$.

\medskip

\noindent\textit{Case 1. There is a school $s^\ast\in\widehat S$ such that
$|\{j\,;\,r_{s^\ast}(j)<r_{s^\ast}(i)\}|<q_{s^\ast}$.}

Consider Step 1 of the first round. We claim that, as long as $i$ has
not yet been assigned, at the beginning of every clinching subround $h$,
\begin{equation}
\label{eq:CT_case1_invariant}
\left|\{j\text{ unassigned}:r_{s^\ast}(j)<r_{s^\ast}(i)\}\right|
<q_{s^\ast}^h.
\end{equation}
The claim holds at the beginning of Step 1 by assumption. Suppose it
holds before some clinching subround $h$, and let $a$ be the number
of unassigned students with higher priority than $i$ at $s^\ast$.
Suppose that $b$ such students and $c$ students with lower priority
than $i$ clinch $s^\ast$ in that subround. Every clinching student
is among the $q_{s^\ast}^h$ highest-priority unassigned students,
so $c\leq q_{s^\ast}^h-a-1$. If $i$ remains unassigned, the
left-hand side after the subround is at most $a-b$, whereas
\begin{equation*}
q_{s^\ast}^{h+1}=q_{s^\ast}^h-b-c>a-b.
\end{equation*}
Students with higher priority than $i$ who are assigned to other
schools only decrease the left-hand side. Thus
\eqref{eq:CT_case1_invariant} holds throughout Step 1 as long as
$i$ remains unassigned. In particular, if $i$ is not assigned in
Step 1, then $s^\ast$ still has positive remaining capacity at the
beginning of Step 2.

If $i$ is assigned in Step 1, she is assigned to her most preferred
available school. Since $s^\ast$ remains available up to the moment at
which $i$ is assigned, the assigned school is weakly preferred by $i$
to $s^\ast$, and therefore belongs to $\widehat{S}$.

Suppose instead that $i$ is not assigned in Step 1. Let the residual
problem at the beginning of Step 2 be denoted by
$(\overline{\succ},\overline{C})$, with remaining students
$\overline{I}$, remaining schools $\overline{S}$, priorities
$\overline{r}$, and remaining numbers of seats $\overline{q}$.
Then $i\in\overline{I}$, $s^\ast\in\overline{S}$, and
$|\{j\,;\,\overline{r}_{s^\ast}(j)
<\overline{r}_{s^\ast}(i)\}|<\overline{q}_{s^\ast}$.

If $i$ is assigned in Step 2, she is assigned to her most preferred
available school. Since $s^\ast$ is still available, the school assigned
to $i$ is weakly preferred to $s^\ast$, and hence belongs to
$\widehat{S}$.

If $i$ is not assigned in Step 2, then the same inequality remains true
for $s^\ast$ in the residual problem at the beginning of the next
round. Indeed, if $s^\ast$ is not in a cycle, then its capacity is not
reduced. If $s^\ast$ is in a cycle, then it points to a highest-priority
remaining student. If it pointed to $i$, then $i$ would be in that
cycle and would be assigned, contrary to the supposition. Hence, when
$s^\ast$ is in a cycle and $i$ is not assigned, $s^\ast$ points to a
student in
$\{j\in\overline I:\overline r_{s^\ast}(j)
<\overline r_{s^\ast}(i)\}$;
that student is removed and $\overline{q}_{s^\ast}$ is reduced by
one. Thus the strict inequality is preserved.

Therefore, if $i$ is not assigned in the first round, then at the
beginning of the second round there is still a school
$s^\ast\in\widehat S$ with positive capacity such that $i$ has fewer
higher-priority competitors at $s^\ast$ than the remaining capacity of
$s^\ast$. This argument does not require $i$ to be eligible to
clinch. Repeating the same argument in each subsequent round shows that $i$ is
eventually assigned either to $s^\ast$ or to a school she prefers to
$s^\ast$. In either case, she is assigned to a school in $\widehat S$.

\medskip

\noindent\textit{Case 2. For every school $s\in\widehat S$,
$|\{j\,;\,r_s(j)<r_s(i)\}|\geq q_s$.}

We show that, after the first round of CT, either $i$ is assigned to a
school in $\widehat S$, or in the residual problem $i$ has a safety-set
contained in $\widehat S$.

If, at some clinching subround of Step 1, there is a school
$s\in\widehat S$ such that
\begin{equation*}
\left|\{j\text{ unassigned}:r_s(j)<r_s(i)\}\right|<q_s^h,
\end{equation*}
then the argument from Case 1 applies from that subround onward.
Therefore, suppose that throughout Step 1, as long as $i$ is not
assigned, for every $s\in\widehat S$ we have
\begin{equation}
\label{eq:CT_case2_ineq}
\left|\{j\text{ unassigned}:r_s(j)<r_s(i)\}\right|\ge q_s^h.
\end{equation}

By the definition of $\Safe^I(\succ_i,\{C\})$ and Hall's theorem,
there is a nonempty high-priority set $A\subseteq\widehat S$ such that
\begin{equation}
\label{eq:CT_hall_initial}
\left|\bigcup_{s\in A}\{j\in I:r_s(j)<r_s(i)\}\right|
<\sum_{s\in A}q_s.
\end{equation}
If $i$ is assigned in Step 1, let $\widetilde\mu$ be the partial
assignment produced before the subround in which she is assigned.
Otherwise, let $\widetilde\mu$ be the partial assignment produced
in all of Step 1. For a set of schools $B$, let $\widetilde\mu(B)$
denote the set of students assigned by $\widetilde\mu$ to schools
in $B$.

By \eqref{eq:CT_case2_ineq}, every student assigned by
$\widetilde\mu$ to a school in $A$ has strictly higher priority
than $i$ at that school. Hence
\begin{equation*}
\widetilde\mu(A)\subseteq
\bigcup_{s\in A}\{j\in I:r_s(j)<r_s(i)\}.
\end{equation*}
Let $(\overline{\succ},\overline{C})$ be the residual problem after
these assignments, with remaining students $\overline I$, remaining
schools $\overline S$, priorities $\overline r$, and remaining
numbers of seats $\overline q$. Let
\begin{equation*}
A^0=\{s\in A:\overline q_s>0\}.
\end{equation*}
Then
\begin{equation*}
\bigcup_{s\in A^0}
\{j\in\overline I:\overline r_s(j)<\overline r_s(i)\}
\subseteq
\left(\bigcup_{s\in A}\{j\in I:r_s(j)<r_s(i)\}\right)
\setminus\widetilde\mu(A).
\end{equation*}
Therefore, using \eqref{eq:CT_hall_initial},
\begin{align}
\left|\bigcup_{s\in A^0}
\{j\in\overline I:\overline r_s(j)<\overline r_s(i)\}\right|
&\leq
\left|\bigcup_{s\in A}\{j\in I:r_s(j)<r_s(i)\}\right|
-|\widetilde\mu(A)|\notag\\
&<\sum_{s\in A}q_s-|\widetilde\mu(A)|
=\sum_{s\in A^0}\overline q_s.
\label{eq:CT_step1_hall}
\end{align}
Thus $A^0$ is a high-priority set for $i$ in the residual problem.
In particular, $A^0$ is nonempty. Hence $i$ has a safety-set in
$(\overline\succ,\overline C)$ contained in $\widehat S$; denote
it by $S^0$.

If $i$ is assigned in Step 1, this residual problem is the one
immediately before the subround in which she is assigned. Since
$S^0$ is nonempty and $i$ receives her most preferred available
school, her assignment belongs to $S^0$, and hence to $\widehat S$.
Otherwise, $(\overline\succ,\overline C)$ is the residual problem
at the beginning of Step 2.

Now consider Step 2. If $i$ is assigned in Step 2, she is assigned to
her most preferred available school. Since $S^0$ is nonempty and
contained in $\widehat S$, the school assigned to $i$ belongs to
$S^0$, and therefore to $\widehat S$.

Suppose instead that $i$ is not assigned in Step 2. Since $S^0$ is a
safety-set for $i$ in $(\overline{\succ},\overline{C})$, there is a
high-priority set $B\subseteq S^0$ such that
\begin{equation}
\label{eq:CT_step2_hall}
\left|\bigcup_{s\in B}
\{j\in\overline I:\overline r_s(j)<\overline r_s(i)\}\right|
<\sum_{s\in B}\overline q_s.
\end{equation}
Let $D\subseteq B$ be the set of schools in $B$ that are in cycles in
Step 2. For each school in $D$, capacity is reduced by one. Moreover,
because $i$ is not assigned in Step 2, no school in $D$ can point to
$i$; otherwise $i$ would be in that cycle. Since each such school points
to its highest-priority remaining student, every school in $D$ points
to a student with higher priority than $i$ at that school. These
pointed-to students are distinct and are removed in Step 2.

Let $(\succ^1,C^1)$ be the residual problem at the beginning of the
second round, with remaining students $I^1$, priorities $r^1$, and
remaining numbers of seats $q^1$. Let
\begin{equation*}
B^1=\{s\in B:q_s^1>0\}.
\end{equation*}
Then
\begin{equation}
\label{eq:CT_next_round_hall}
\left|
\bigcup_{s\in B^1}
\{j\in I^1:r_s^1(j)<r_s^1(i)\}
\right|
<
\sum_{s\in B^1}q_s^1.
\end{equation}
Thus $B^1$ is a high-priority set for $i$ in the residual problem at
the beginning of the second round. Hence $i$ has a safety-set in
$(\succ^1,C^1)$ contained in $S^0$, and therefore contained in
$\widehat S$.

The same argument can now be repeated round by round. At the beginning
of each round in which $i$ is still unassigned, either the Case 1
argument applies, or the Case 2 argument shows that $i$ has a safety-set
contained in her previous safety-set. Since the set of students is finite
and at least one student is assigned in every round, $i$ is eventually
assigned, and the assignment must be to a school in
$\widehat S=\Safe^I(\succ_i,\{C\})$.
\end{proof}

\subsection{First Clinch and Trade mechanism}
\label{sec:first-clinch-trade}

The First Clinch and Trade mechanism works as follows. At each round,
unassigned students point to their most preferred available school
and each available real school points to the highest-priority
unassigned student. The assignment in each round is obtained in
two steps.

\begin{itemize}
\item[\textbf{Step 1}] If a student $i$ points to a real school $s$
  and was among the $q_s(C_s)$ highest-priority students at $s$
  in the original problem, assign $i$ to $s$ (i.e., $i$ clinches
  $s$). Thus the clinching condition is
  \begin{equation*}
  \left|\{j\in I:r_s(j)<r_s(i)\}\right|<q_s(C_s).
  \end{equation*}
  This condition is fixed throughout the algorithm. Every student
  pointing to $s_0$ is assigned to $s_0$. All these assignments
  are determined simultaneously. Remove the assigned students
  and reduce each real school's remaining capacity by the number
  of students who clinched that school.

\item[\textbf{Step 2}] Keep the pointers set at the beginning of
  the round. For each cycle consisting of remaining students and
  available real schools, assign each student in the cycle to
  the school she points to. Remove all students assigned in these
  cycles and reduce the capacity of each school in a cycle by one.
\end{itemize}

Repeat these two steps, setting new pointers at the beginning of
each round, until all students are assigned.

\begin{lemma}\label{lemma:clinch-trade-first_clinch-trade_incontestable}
The First Clinch and Trade mechanism respects safety-sets.
\end{lemma}

\begin{proof}
Fix a problem $(\succ,C)$ and a student $i$. Let
\begin{equation*}
\widehat S=\Safe^I(\succ_i,\{C\}).
\end{equation*}
We show that FCT assigns $i$ to a school in $\widehat S$.
If $s_0\in\widehat S$, individual rationality of FCT and the fact
that $\widehat S$ is an upper contour set give the desired
conclusion. Hence suppose that $s_0\notin\widehat S$.

We first show that every student who clinches a real school $s$
is among its highest-priority unassigned students up to its
remaining number of seats. For each real school $s$, let
\begin{equation*}
E_s=\left\{j\in I:
\left|\{h\in I:r_s(h)<r_s(j)\}\right|<q_s(C_s)\right\}.
\end{equation*}
Thus $E_s$ is the set of students who can clinch $s$ under FCT.
At any point, let $\overline I$ denote the remaining students and
$\overline q_s$ the remaining number of seats at $s$. We claim that
\begin{equation}
\label{eq:FCT_initial_eligibility_invariant}
|E_s\cap\overline I|\leq\overline q_s.
\end{equation}
The claim holds initially. A clinch at $s$ removes one member of
$E_s$ and one seat. Assignments elsewhere can only reduce the
left-hand side. If a trading cycle uses $s$ and a member of
$E_s$ remains after clinching, then the student to whom $s$
points belongs to $E_s$. Indeed, she had highest priority when
the pointers were set, and every member of $E_s$ has higher
priority than every student outside $E_s$. Since $s$ is in an
executed cycle, its pointed-to student has not clinched and is
removed in that cycle. Hence the loss of one seat is accompanied
by the removal of at least one member of $E_s$. If no member
of $E_s$ remains, the inequality holds trivially. This proves
\eqref{eq:FCT_initial_eligibility_invariant} throughout the algorithm.

Every student with higher priority than a member $j$ of $E_s$
also belongs to $E_s$. Therefore, whenever $j$ clinches $s$,
\begin{equation*}
\left|\{h\in\overline I:r_s(h)<r_s(j)\}\right|
\leq |E_s\cap\overline I|-1<\overline q_s.
\end{equation*}
Thus every FCT clinch satisfies the priority condition used in
the proof of \thref{lemma:CT-incontestable}.

The proof follows the same invariant as in the proof of
\thref{lemma:CT-incontestable}. At the beginning of any round in which
$i$ is still unassigned, suppose that $i$ has a safety-set $S^k$ in the
current residual problem, with $S^k\subseteq\widehat S$. This
holds initially with $S^0=\widehat S$.

Consider the clinching step of that round. Since $S^k$ is nonempty
and excludes $s_0$, student $i$ does not point to $s_0$.
If $i$ clinches a school, then she clinches her most preferred
available school. Since $S^k$ is a safety-set and is nonempty,
the school clinched by $i$ belongs to $S^k$, and hence to
$\widehat S$.

If $i$ does not clinch a school, then the same counting arguments
used for Step 1 in Cases 1 and 2 of the proof of
\thref{lemma:CT-incontestable} show that,
after removing the students who clinch and reducing capacities
accordingly, $i$ has a safety-set $\widetilde S^k$ in the residual problem
such that
\begin{equation*}
\widetilde S^k\subseteq S^k\subseteq \widehat S.
\end{equation*}
Now consider the trading step. Although the pointers were set before
clinching, a school in an executed cycle points to a student who
remains unassigned and therefore still has highest priority among
the remaining students. Likewise, a student in an executed cycle
points to a school that is still available and remains her most
preferred available school. Thus the same cycle-counting argument
applies to these cycles.

If $i$ is assigned in a cycle, then she is assigned to her most
preferred available school, which belongs to $\widetilde S^k$,
and hence to $\widehat S$. If $i$ is not assigned in
the trading step, then the same cycle-counting argument used for Step 2
in the proof of \thref{lemma:CT-incontestable} shows that, at the
beginning of the next round, $i$ has a safety-set $S^{k+1}$ satisfying
\begin{equation*}
S^{k+1}\subseteq \widetilde S^k\subseteq S^k\subseteq\widehat S.
\end{equation*}

By induction, as long as $i$ is not assigned, she continues to have a
safety-set contained in $\widehat S$. Every round assigns at least
one student: if no student is assigned in the clinching step, no
pointer is removed and there is a trading cycle. Since the set of
students is finite, $i$ is eventually assigned, and the assigned
school must belong to
$\widehat S=\Safe^I(\succ_i,\{C\})$.
\end{proof}

\section{Group protection}
\label{sec:group-protection}

\subsection{About group protection}
\label{sec:necess-group-prot}

The following example shows that Theorem~\ref{thm:attainability} need not
hold without group protection, even when schools have responsive
priorities.

There are four schools, $s_0$, $s_1$, $s_2$, and $s_3$, and four
students, $i_h$, $h=1,\ldots,4$. Let $C^0$ be the choice profile
induced by the priority orderings in
Table~\ref{tab:incontestable_not_group_protection} and one seat at
each real school. Thus $q_s(C_s^0)=1$ for $s=s_1,s_2,s_3$,
while $C_{s_0}^0(J)=J$ and $q_{s_0}(C_{s_0}^0)=4$.
\begin{table}[ht!]
  \centering
  \begin{tabular}{ccc}
    $r_{s_1}$ & $r_{s_2}$ & $r_{s_3}$\\
    \hline
    $i_2$ & $i_2$ & $i_2$ \\
    $i_1$ & $i_3$ & $i_4$ \\
    $i_3$ & $i_1$ & $i_1$ \\
    $i_4$ & $i_4$ & $i_3$
  \end{tabular}
  \caption{Priorities in the example without group protection}
  \label{tab:incontestable_not_group_protection}
\end{table}

Consider the domain of admissible path-independent choice profiles
on these students and schools. Define a mechanism $\varphi$ as follows.
Whenever $C=C^0$, use a serial dictatorship with the fixed order
$i_2,i_1,i_3,i_4$: each student chooses her most preferred school
among those with a remaining seat, including $s_0$, which is always
available. Whenever $C\neq C^0$, use the student-optimal stable
mechanism.

When $C\neq C^0$, the mechanism is stable, so it is non-wasteful
and respects safety-sets. When $C=C^0$, it is also non-wasteful:
if a student prefers a school to her assignment, that school was
already full when she chose. We now check that it also respects
safety-sets at $C^0$, for every preference profile.

Student $i_2$ has highest priority at every real school. Hence her
safety-set is the singleton containing her most preferred school,
which she receives under $\varphi$. For each of the other students,
all high-priority sets contain $s_0$. Indeed, filling
$(s_1,s_2,s_3)$ with $(i_2,i_3,i_4)$, $(i_1,i_2,i_4)$, and
$(i_1,i_3,i_2)$, respectively, and assigning the remaining student
to $s_0$, rejects $i_1$, $i_3$, and $i_4$, respectively, at every
real school. Therefore, for every preference relation $\succ_i$,
\begin{equation*}
\Safe^I(\succ_i,\{C^0\})
=\{s\in S:s\succeq_i s_0\}
\qquad\text{for }i=i_1,i_3,i_4.
\end{equation*}
Since each of these students chooses from a set containing $s_0$,
her assignment belongs to her safety-set. Thus $\varphi$ respects
safety-sets throughout its domain and is $\Pi^C$-incontestable.

We now show that the conclusion of Theorem~\ref{thm:attainability}
fails. Let $\mathcal K_{i_1}=\{C^0\}$, so that $i_1$ knows the
schools' choice functions perfectly, and let her preferences be
\begin{equation*}
s_1\succ_{i_1}s_2\succ_{i_1}s_3\succ_{i_1}s_0.
\end{equation*}
Then
\begin{equation*}
\Safe^I(\succ_{i_1},\mathcal K_{i_1})
=\{s_0,s_1,s_2,s_3\}.
\end{equation*}
For every preference profile $\succ_{-i_1}$, student $i_2$ chooses
first and student $i_1$ chooses second. If $i_2$ chooses $s_1$,
then $i_1$ receives $s_2$; otherwise, $i_1$ receives $s_1$.
Both cases can occur. Thus the set of attainable schools for
$i_1$ is exactly $\{s_1,s_2\}$. In particular, $s_3$ belongs
to her safety-set but is not attainable.

It remains to note that $\varphi$ violates group protection.
Keep $i_1$'s preferences as above and let
\begin{equation*}
\begin{aligned}
s_1&\succ_{i_2}s_2\succ_{i_2}s_3\succ_{i_2}s_0,\\
s_2&\succ_{i_3}s_1\succ_{i_3}s_3\succ_{i_3}s_0,\\
s_3&\succ_{i_4}s_1\succ_{i_4}s_2\succ_{i_4}s_0.
\end{aligned}
\end{equation*}
Let $T=\{s_1,s_2\}$, $G_{s_1}=\{i_2\}$, and
$G_{s_2}=\{i_3\}$. Then $G=\{i_2,i_3\}$ and
$Z=\{i_1,i_4\}$. Each designated student ranks her school
first, each group fills its school's capacity, and
\begin{equation*}
C_{s_1}^0(G_{s_1}\cup Z)=G_{s_1},
\qquad
C_{s_2}^0(G_{s_2}\cup Z)=G_{s_2}.
\end{equation*}
However, the serial dictatorship gives
\begin{equation*}
\bigl(\varphi_{i_1}(\succ,C^0),\varphi_{i_2}(\succ,C^0),
\varphi_{i_3}(\succ,C^0),\varphi_{i_4}(\succ,C^0)\bigr)
=(s_2,s_1,s_3,s_0).
\end{equation*}
Consequently,
\begin{equation*}
\varphi_{s_1}(\succ,C^0)\cup\varphi_{s_2}(\succ,C^0)
=\{i_2,i_1\}\neq G.
\end{equation*}
Thus $\varphi$ is $\Pi^C$-incontestable and non-wasteful, but
does not satisfy group protection, and the conclusion of
Theorem~\ref{thm:attainability} fails.

\subsection{Stable mechanisms}
\label{sec:stable-mechanisms}

\begin{lemma}\label{lem:group-protection-stable-pareto}
Suppose that schools' choice functions are path independent.
Every stable mechanism satisfies group protection, as does every
mechanism that weakly Pareto improves on the outcome of
student-proposing Deferred Acceptance at each problem.
\end{lemma}

\begin{proof}
Fix a problem $(\succ,C)$, a stable assignment
$\mu\in\mathcal M(C)$, and sets $T$, $(G_s)_{s\in T}$, $G$, and
$Z$ satisfying the conditions of group protection.
Fix $s\in T$, and let $M=\mu(s)$.
For every $j\in G_s$, we have $C_s(M\cup\{j\})=M$.
Indeed, if $j\in M$, this follows from feasibility and the
no-unacceptability assumption. If $j\notin M$, then $j$ ranks
$s$ first, so stability implies $j\notin C_s(M\cup\{j\})$;
consistency then gives $C_s(M\cup\{j\})=C_s(M)=M$.
Applying path independence successively yields
\begin{equation*}
C_s(M\cup G_s)=M.
\end{equation*}
Suppose that $z\in M\cap Z$. Since $C_s(G_s\cup Z)=G_s$,
consistency implies $C_s(G_s\cup\{z\})=G_s$.
By substitutability, $z$ is also rejected from $M\cup G_s$,
contradicting $C_s(M\cup G_s)=M$.
Thus $\mu(s)\subseteq G$ for every $s\in T$.

Stable assignments are non-wasteful. If some school $s\in T$
had a vacancy, then, since $|G_s|=q_s(C_s)$, some student
$j\in G_s$ would not be assigned to $s$. That student ranks $s$
first, contradicting non-wastefulness. Hence every school in $T$
is full. Since $|G|=\sum_{s\in T}q_s(C_s)$ and
$\mu(T)\subseteq G$, we obtain $\mu(T)=G$.

We now prove the second claim. Let
$\mu^D=\mathrm{SOSM}(\succ,C)$ be the outcome of student-proposing
Deferred Acceptance. We first show that $\mu^D(s)=G_s$ for every
$s\in T$. Every student in $G_s$ proposes to $s$ in the first
round. Suppose that a member of $G$ is rejected, and consider the
first round in which this occurs. Until that round, no member of
$G_s$ has left $s$, and no member of $G\setminus G_s$ has proposed
to $s$. Thus, at any school $s\in T$, the set from which the school
chooses at that round is $G_s\cup Z_s$ for some $Z_s\subseteq Z$.
Since $C_s(G_s\cup Z)=G_s$, consistency gives
\begin{equation*}
C_s(G_s\cup Z_s)=G_s.
\end{equation*}
Therefore no member of $G_s$ is rejected, a contradiction.
Hence $\mu^D(s)=G_s$ for every $s\in T$.

Now let $\mu'\in\mathcal M(C)$ weakly Pareto improve on $\mu^D$.
Every member of $G_s$ receives her first choice $s$ under $\mu^D$,
so she must still receive $s$ under $\mu'$. Feasibility and
$|G_s|=q_s(C_s)$ imply $\mu'(s)=G_s$ for every $s\in T$.
Thus $\mu'(T)=G$, which proves the second claim.
\end{proof}

\subsection{Boston}
\label{sec:boston}

Under responsive priorities, every member of $G_s$ applies to $s$ in
the first round of Boston. Members of other designated groups apply to
their own schools. Since $|G_s|=q_s$ and every member of $G_s$ has
higher priority at $s$ than every outsider, school $s$ permanently
accepts exactly $G_s$. Thus Boston satisfies group protection.

\subsection{TTC mechanism}
\label{sec:ttc-mechanism}

\begin{lemma}\label{lem:TTC-group-protection}
Under responsive priorities, TTC satisfies group protection.
\end{lemma}

\begin{proof}
Fix a problem $(\succ,C)$ and sets $T$, $(G_s)_{s\in T}$, $G$,
and $Z$ satisfying the conditions of group protection.
Since $|G_s|=q_s(C_s)$ and $C_s(G_s\cup Z)=G_s$, every student
in $G_s$ has higher priority at $s$ than every student in $Z$.
We show that TTC assigns every member of $G_s$ to $s$.

Initially, for every $s\in T$, the number of unassigned students
in $G_s$ equals the number of available seats at $s$.
Suppose this equality holds at the beginning of a round, and every
previously assigned member of $G_s$ has received $s$.
Every remaining member of $G_s$ points to $s$, her first choice,
which still has an available seat. Every school in $T$ with an
available seat points to a member of $G$: one of its own designated
students remains and has higher priority than every outsider.
Thus any cycle containing such a school or a member of $G$ stays
within the remaining schools in $T$ and students in $G$.
Every student assigned in such a cycle receives her designated
school. Each participating school loses one seat and one of its
designated students. Cycles outside these sets change neither
quantity. The induction hypothesis is therefore preserved.
Eventually every member of $G_s$ is assigned to $s$.
Feasibility gives $\mathrm{TTC}_s(\succ,C)=G_s\subseteq G$ for
every $s\in T$, as required.
\end{proof}

\subsection{Efficiency Adjusted Deferred Acceptance mechanism}
\label{sec:effic-adjust-deferr}

We consider the simplified version of the Efficiency Adjusted Deferred
Acceptance mechanism (SEADAM) proposed by \citet{tang2014new}, when
all students consent. Under
responsive priorities, SEADAM selects the same assignment as EADAM
with full consent, as
well as the optimal legal mechanism \citep{ehlers2020legal} and the
priority efficient mechanism \citep{reny2022efficient}.

Fix a responsive-priority problem $(\succ,C)$, and let $(r,q)$ be the
priority and capacity representation of $C$. For an assignment $\mu$,
we say that a school $s$ is \emph{under-demanded at $\mu$} if there is
no student $i$ such that
\begin{equation*}
s\succ_i \mu(i).
\end{equation*}

SEADAM is a multi-round algorithm. Let
\begin{equation*}
(\succ^1,C^1)=(\succ,C).
\end{equation*}
For each round $k\ge 1$, let
\begin{equation*}
\mu^k=\mathrm{SOSM}(\succ^k,C^k)
\end{equation*}
be the student-optimal stable assignment for the current problem; that
is, the outcome of student-proposing Deferred Acceptance. Let $I^k$
and $S^k$ denote the remaining students and schools, with $s_0\in S^k$.
Let $D^k$ be the set of schools in $S^k$, including $s_0$, that are
under-demanded at $\mu^k$, and let
\begin{equation*}
F^k=\{j\in I^k:\mu^k(j)\in D^k\}.
\end{equation*}
If $F^k=\varnothing$, the algorithm stops and assigns every remaining
student according to $\mu^k$, together with the assignments finalized
in earlier rounds. If $F^k\neq\varnothing$, every student in $F^k$
is finalized at her current assignment. If all remaining students
have been finalized, the algorithm stops. Otherwise, it moves to
round $k+1$ by removing the finalized students and all real schools
in $D^k$. Preferences and priorities are restricted to the remaining
students and schools, and each remaining real school keeps its
unallocated seats. Choice functions are the responsive rules for
those priorities and seats. The null school remains available to
every remaining student, with choice rule
$C_{s_0}^{k+1}(J)=J$. Denote the resulting problem by
$(\succ^{k+1},C^{k+1})$.

It is well known that EADAM weakly Pareto dominates the
student-optimal stable mechanism. Hence,
Lemma~\ref{lem:group-protection-stable-pareto} implies
that it satisfies group protection. This argument applies with
partial as well as full consent. The same conclusion applies to
the other mechanisms just mentioned whenever they weakly Pareto
improve on the student-optimal stable assignment.

\subsection{Clinch and Trade, and First Clinch and Trade}
\label{sec:clinch-trade-first}

\begin{lemma}\label{lem:CT-FCT-group-protection}
Under responsive priorities, the Clinch and Trade mechanism and the
First Clinch and Trade mechanism satisfy group protection.
\end{lemma}

\begin{proof}
Fix a problem $(\succ,C)$ and sets $T$, $(G_s)_{s\in T}$, $G$,
and $Z$ satisfying the conditions of group protection.
Since $|G_s|=q_s(C_s)$ and $C_s(G_s\cup Z)=G_s$, every student
in $G_s$ has higher priority at $s$ than every student in $Z$.
We show that each mechanism assigns every member of $G_s$ to $s$.

Consider either algorithm. At any point, let $\overline I$ be the
set of unassigned students and let $\overline q_s$ be the number of
available seats at $s$. We show by induction over the clinching
and trading steps that every previously assigned member of $G_s$
has received $s$, and that
\begin{equation}\label{eq:CT-FCT-group-invariant}
|G_s\cap\overline I|=\overline q_s
\qquad\text{for every }s\in T.
\end{equation}
Both claims hold initially. Suppose they hold before the next
step. Every remaining member of $G_s$ still has her first-choice
school $s$ available, and therefore can only clinch or point to $s$.

\medskip
\noindent\textit{Clinching steps.}
Under CT, a student can clinch $s$ only if she is among the
$\overline q_s$ highest-priority unassigned students at $s$.
Every remaining student in $Z$ has at least
$|G_s\cap\overline I|=\overline q_s$ higher-priority unassigned
students at $s$. Thus no student in $Z$ can clinch a school in $T$.

Under FCT, a student can clinch $s$ only if she was among the
$q_s(C_s)$ highest-priority students at $s$ in the original problem.
Every student in $Z$ initially has all $q_s(C_s)$ members of $G_s$
ahead of her at $s$. Thus no student in $Z$ can clinch a school
in $T$ under FCT either.

Under either mechanism, therefore, a student who clinches a school
$s\in T$ must belong to $G_s$. Every such clinch removes one
member of $G_s$ and one seat at $s$. Clinches at other schools
remove no member of $G$. Hence both claims are preserved after
each clinching step, including when several clinches are performed
simultaneously.

\medskip
\noindent\textit{Trading steps.}
Whenever a school $s\in T$ with an available seat sets its
pointer, it points to a member of $G$: by
Eq.~\eqref{eq:CT-FCT-group-invariant}, one of its own designated
students remains and has higher priority than every outsider.
Every remaining member of $G_s$ points to $s$.
If pointers are set before clinching, a pointer to a student who
subsequently clinches cannot belong to an executed trading cycle.
Consequently, every executed cycle containing a school in $T$ or
a member of $G$ stays within the remaining schools in $T$ and
students in $G$.

Every student assigned in such a cycle receives her designated
school. Each participating school loses one seat and one of its
designated students. Cycles outside these sets change neither
quantity. Thus both claims are preserved after each trading step.

The induction applies throughout both algorithms. Neither algorithm
can terminate with a member of $G_s$ unassigned, since
Eq.~\eqref{eq:CT-FCT-group-invariant} would imply that $s$ still
has an available seat. Hence every member of $G_s$ is assigned to
$s$. Feasibility and $|G_s|=q_s(C_s)$ give
\begin{equation*}
\mathrm{CT}_s(\succ,C)=\mathrm{FCT}_s(\succ,C)=G_s
\qquad\text{for every }s\in T.
\end{equation*}
Taking the union over $s\in T$ proves group protection for both
mechanisms.
\end{proof}

\subsection{Group protection v. stable dominating}
\label{sec:group-protection-v}

Stable domination and group protection are logically independent,
even when schools have responsive priorities and unit capacities.
The distinction is that a stable-dominating assignment need only
Pareto dominate \emph{some} stable assignment.

Throughout this section, schools have responsive
priorities.

\begin{definition}[\cite{troyan2020obvious}]
\label{def:stable-domination}
An assignment $\mu$ is
\emph{stable dominating} if it is stable or Pareto dominates some
stable assignment $\nu$: $\mu(i)\succeq_i\nu(i)$ for every
$i\in I$, with a strict preference for at least one student. A
mechanism is stable dominating if it selects a stable-dominating
assignment at every problem.
\end{definition}

The following example shows that stable domination does not imply
group protection.

\begin{example}
\label{ex:stable-domination-not-group-protection}
Let $I=\{i,j,k\}$ and $S=\{s_0,s_1,s_2,s_3\}$. Each non-null
school has one seat. Preferences and priorities are given in
Table~\ref{tab:stable-domination-counterexample}; entries are
ordered from highest to lowest, and every student ranks $s_0$ last.

\begin{table}[htbp]
\centering
\begin{tabular}{ccccccc}
$\succ_i$ & $\succ_j$ & $\succ_k$ &\qquad &
$C_{s_1}$ & $C_{s_2}$ & $C_{s_3}$ \\
\cline{1-3}\cline{5-7}
$s_1$ & $s_2$ & $s_1$ & \qquad &$j$ & $i$ & $k$ \\
$s_3$ & $s_1$ & $s_3$ & \qquad &$i$ & $j$ & $i$ \\
$s_2$ & $s_3$ & $s_2$ &\qquad & $k$ & $k$ & $j$ \\
\end{tabular}
\caption{Stable domination need not imply group protection.}
\label{tab:stable-domination-counterexample}
\end{table}

Consider the assignments
\begin{align*}
\bigl(\nu(i),\nu(j),\nu(k)\bigr)
  &= (s_2,s_1,s_3), \\
\bigl(\mu(i),\mu(j),\mu(k)\bigr)
  &= (s_3,s_2,s_1).
\end{align*}
The assignment $\nu$ is stable, since each school holds its
highest-priority student. Every student strictly prefers $\mu$
to $\nu$, so $\mu$ is stable dominating.

Now take $T=\{s_1,s_2\}$, $G_{s_1}=\{i\}$, and
$G_{s_2}=\{j\}$. Then $G=\{i,j\}$ and $Z=\{k\}$. Students $i$
and $j$ rank $s_1$ and $s_2$ first, respectively, both groups fill
their school's capacity, and
\begin{equation*}
C_{s_1}(\{i,k\})=\{i\},
\qquad
C_{s_2}(\{j,k\})=\{j\}.
\end{equation*}
All the premises of group protection hold. However,
\begin{equation*}
\mu(s_1)\cup\mu(s_2)=\{k,j\}\neq\{i,j\}=G.
\end{equation*}
Thus, selecting $\mu$ violates group protection.\footnote{
To obtain a mechanism counterexample, select $\mu$ at this problem
and the student-proposing deferred acceptance assignment at every
other problem. The resulting mechanism is stable dominating and
violates group protection.}
\end{example}

Conversely, TTC satisfies group protection but need not be stable
dominating. Consider instead the preferences and priorities in
Table~\ref{tab:ttc-counterexample-short}, with the same students,
schools, and capacities as in
Example~\ref{ex:stable-domination-not-group-protection}. Every
student again ranks $s_0$ last.

\begin{table}[htbp]
\centering
\begin{tabular}{ccccccc}
$\succ_i$ & $\succ_j$ & $\succ_k$ &\qquad &
$C_{s_1}$ & $C_{s_2}$ & $C_{s_3}$ \\
\cline{1-3}\cline{5-7}
$s_1$ & $s_1$ & $s_2$ & \qquad &$k$ & $i$ & $i$ \\
$s_2$ & $s_2$ & $s_3$ &\qquad & $i$ & $j$ & $k$ \\
$s_3$ & $s_3$ & $s_1$ & \qquad &$j$ & $k$ & $j$ \\
\end{tabular}
\caption{TTC need not select a stable-dominating assignment.}
\label{tab:ttc-counterexample-short}
\end{table}

The  unique stable assignment is $\nu=(s_1,s_2,s_3)$, and the
assignment obtained with TTC is $\mu'=(s_1,s_3,s_2)$. Since
$\nu(j)=s_2\succ_j s_3=\mu'(j)$, the TTC  is not stable
dominating.

\section{Other information classes}
\label{sec:proofs-prop-sect}

\begin{proposition}\label{prop:pi-mu}
Suppose that, at each real school $s$, $\mathcal C_s$ contains every
responsive choice function with capacity in $\{1,\ldots,|I|\}$.
A mechanism is $\Pi^\mu$-incontestable if and only if it is
individually rational and, for every problem $(\succ,C)$, every
student $i$, and every real school $s$,
\begin{equation*}
s\succ_i\varphi_i(\succ,C)
\quad\Longrightarrow\quad
\varphi_s(\succ,C)\neq\varnothing.
\end{equation*}
\end{proposition}

\begin{proof}
Suppose first that $\varphi$ is individually rational and never
leaves a preferred real school empty.
Fix $(\succ,C)$, let $\mu=\varphi(\succ,C)$, and fix a student
$i$ and a rule in $\Pi^\mu$.
For every real school $s\succ_i\mu(i)$, we have $|\mu(s)|\geq 1$.
Choose a responsive rule $\widehat C_s$ with capacity
$q_s(\widehat C_s)=|\mu(s)|$ that ranks all of $\mu(s)$ above $i$.
Then $i\notin\widehat C_s(\mu(s)\cup\{i\})$.
For every other school, retain its original choice function.
The domain assumption implies $\widehat C\in\mathcal C$, and
$\mu\in\mathcal M(\widehat C)$.
Since the information rule imposes no choice-profile restrictions beyond
the maintained domain and compatibility with the assignment-side
information, and since $\mu$ is compatible with its 
assignment-side information, $(\widehat C,\mu)$ is compatible.
Individual rationality rules out a block with the null school.
Thus, in this state $i$ has no blocking school, and she has no
legitimate complaint.

Conversely, if $\varphi$ is not individually rational, at some
outcome $\mu$ a student $i$ prefers $s_0$ to $\mu(i)$.
The null school chooses $i$ in every compatible state, so $i$ has
a legitimate complaint.
If instead some student $i$ prefers an empty real school $s$ to
her assignment, use the rule in $\Pi^\mu$ under which the entire
realized assignment $\mu$ is admissible information. 
For every compatible choice profile $\widehat C$, positivity of
$q_s(\widehat C_s)$ and the no-unacceptability assumption give
\begin{equation*}
i\in\widehat C_s(\mu(s)\cup\{i\})
=\widehat C_s(\{i\}).
\end{equation*}
Thus $i$ has a legitimate complaint. This proves necessity.
\end{proof}

\begin{proposition}\label{prop:pi-cmu-stability}
A mechanism $\varphi$ is $\Pi^{C,\mu}$-incontestable if and only if
$\varphi(\succ,C)$ is stable for every problem $(\succ,C)$.
\end{proposition}

\begin{proof}
Let $\mu=\varphi(\succ,C)$.

We first show the only-if direction. Suppose that $\varphi$ is
$\Pi^{C,\mu}$-incontestable. Consider the complete-revelation rule
$\pi^{F}\in\Pi^{C,\mu}$ defined as follows: for every problem
$(\succ,C)$, assignment $\mu$, and student $i$,
\begin{equation*}
\pi_i^{F,E}(\succ,C,\mu)
=
\{C\}\times \{\mu\}.
\end{equation*}
Thus, ex post, the entire realized choice profile and assignment are
admissible information for student $i$.

Suppose, toward a contradiction, that $\varphi(\succ,C)=\mu$ is not
stable for some problem $(\succ,C)$. Then there exist a student $i$ and
a school $s$ such that
\begin{equation*}
s\succ_i \mu(i)
\qquad\text{and}\qquad
i\in C_s(\mu(s)\cup\{i\}).
\end{equation*}
Under this complete-revelation rule, the only compatible pair
$(\widehat C,\widehat\mu)$ is $(C,\mu)$. Hence, for every compatible
$(\widehat C,\widehat\mu)$, there exists a school, namely $s$, such
that
\begin{equation*}
s\succ_i \mu(i)
\qquad\text{and}\qquad
i\in \widehat C_s(\widehat\mu(s)\cup\{i\}).
\end{equation*}
Therefore student $i$ has a legitimate complaint at $\mu$ under
$\pi^F$. This contradicts $\Pi^{C,\mu}$-incontestability of
$\varphi$. Thus $\varphi(\succ,C)$ must be stable for every problem
$(\succ,C)$.

We now prove the if direction. Suppose that $\varphi(\succ,C)$ is
stable for every problem $(\succ,C)$. Let $\pi\in\Pi^{C,\mu}$ be any
information rule, let $(\succ,C)$ be any problem, and let
\begin{equation*}
\mu=\varphi(\succ,C).
\end{equation*}
Fix a student $i$, and let
\begin{equation*}
\Hi=\pi_i^E(\succ,C,\mu)
\end{equation*}
be her ex-post information set. By correctness of information,
$(C,\mu)\in\Hi$, and therefore the true pair $(C,\mu)$ is
compatible with $\Hi$.

Suppose, toward a contradiction, that student $i$ has a legitimate
complaint at $\mu$ given $\Hi$. Then, by definition, for every pair
$(\widehat C,\widehat\mu)$ compatible with $\Hi$, there exists a school
$s$ such that
\begin{equation*}
s\succ_i \mu(i)
\qquad\text{and}\qquad
i\in \widehat C_s(\widehat\mu(s)\cup\{i\}).
\end{equation*}
Since the true pair $(C,\mu)$ is compatible with $\Hi$, this condition
must hold for $(\widehat C,\widehat\mu)=(C,\mu)$. Hence there exists
$s\in S$ such that
\begin{equation*}
s\succ_i \mu(i)
\qquad\text{and}\qquad
i\in C_s(\mu(s)\cup\{i\}),
\end{equation*}
which means that $(i,s)$ blocks $\mu$ at the problem $(\succ,C)$.
This contradicts the stability of $\mu=\varphi(\succ,C)$.

Thus no student has a legitimate complaint under any
$\pi\in\Pi^{C,\mu}$, at any problem. Therefore $\varphi$ is
$\Pi^{C,\mu}$-incontestable.
\end{proof}

\section{$\Pi^{C}$-Contestable mechanisms}
\label{sec:cont-mech}

\subsection{The Equitable Top Trading Cycles mechanism}
\label{sec:equit-top-trad}

The Equitable Top-Trading Cycles mechanism (ETTC), proposed by
\cite{hakimov2018equitable},  is a variant of the
Top Trading Cycles mechanism that aims at producing efficient assignment in the spirit of the TTC
mechanism but that can generate fewer instances of justified
envy. Like TTC, ETTC is efficient and group strategyproof.
We give \posscite{hakimov2018equitable} description of ETTC below. At
each step, students are first pre-assigned a school (the 
\textit{inheritance} part in the steps of the algorithm). Then,
students point to student--school pairs. 
In each step, a student's best choice is her most-preferred school
among those appearing in a remaining student--school pair after the
inheritance part of that step. All pairs considered in the pointing
part are remaining pairs.

\textbf{Step} $\mathbf{k}$, $\mathbf{k\geq 1}$
\begin{itemize}
\item[]\textbf{Inheritance}

  If $k= 1$, then all slots are available
  for inheritance and assigned to students one by one following
  schools’ priority orders to form student–school pairs. Otherwise,
  for each school $s$ such that \textit{(i)} there are slots of school $s$ that
  remained to be inherited from previous steps, and \textit{(ii)} there are no
  existing pairs that were assigned to a slot of school s at a previous
  steps of the algorithm, i.e., the market has completely run out of
  the slots of school $s$, the slots that remained to be inherited are
  assigned to the remaining students one by one following the priority
  order for school $s$ to form new student–school pairs.

\item[]\textbf{Pointing:}

  Each
  student–school pair $(i,s)$ points to the student–school pair $(i' ,s')$
  such that \textit{(i)} school $s'$ is the best choice of student $i$ and \textit{(ii)}
  student $i'$ is the student with the highest priority for school $s$
  among the students who are assigned to a slot from school $s'$. If
  student $i$ is already assigned to one slot from his best choice
  school, then all student–school pairs containing him point to that
  student–school pair.

\item[]\textbf{Trading:}

  There is at least one cycle. In each
  cycle, corresponding trades are performed, and all student–school
  pairs that participate in a cycle are removed. It is possible that
  student–school pairs containing the same student ---say, student
  $i$---appear in the same cycle or in different cycles. In such a
  case, student $i$ is placed in 
his best choice, and the other slots of that school (to which the student–school
pairs containing him are pointing in those other cycles) remain to be inherited.
For each student–school pair $(i,s)$ that participates in a cycle, the slots assigned
to student $i$ in other student–school pairs that do not participate in a cycle also
remain to be inherited.
\end{itemize}

However,
ETTC is a $\Pi^{C}$-contestable mechanism. To see this, 
consider the example depicted in Table~\ref{tab:ETTC}, where schools
$s_1$ and $s_2$ have one seat and school $s_3$ has two seats.

\begin{table}[ht!]
	\centering
	\begin{tabular}{cccccccc}
		$i_1$ & $i_2$ & $i_3$  & $i_4$ &\qquad  & $s_1$ & $s_2$ & $s_3$\\
		\cline{1-4}\cline{6-8}
		$s_1$ & $s_2$ &  $s_3$ & $s_1$ &  & $i_3$ & $i_3$ & $i_1$\\
		&  &  &  $s_2$ &  & $i_4$ & $i_4$ & $i_2$\\
		&  &  &  $s_3$ &  & $i_1$ & $i_2$ & $i_3$\\
		&  &  &   &  & $i_2$ & $i_1$ & $i_4$\\
	\end{tabular}
	\caption{The Equitable Top Trading Cycles mechanism is $\Pi^{C}$-contestable}
	\label{tab:ETTC}
\end{table}

In the ETTC mechanism, seats at schools are first pre-assigned to the
students, one by one, starting with the student with the highest
priority ranking, until all seats have been pre-assigned. In our example
student $i_3$ is pre-assigned the seats of schools $s_1$ and $s_2$, and
students $i_1$ and $i_2$ are pre-assigned the two seats at school $s_3$.

So, the pair $(i_1,s_3)$ points to the pair $(i_3,s_1)$ and the pair
$(i_2,s_3)$ points to the pair $(i_3,s_2)$. Student $i_3$'s most
preferred school is $s_3$.
The two seats at $s_3$ are
pre-assigned to $i_1$ and $i_2$. 
At school $s_1$, student $i_1$ has
higher priority than $i_2$, so $(i_3 ,s_1)$ points to $(i_1,s_3)$.
Similarly, $(i_3,s_2)$ points to $(i_2,s_3)$ because at $s_2$ student
$i_2$ has a higher priority than $i_1$. 

We then have two cycles involving $(i_3,s_1)$ and $(i_1,s_3)$, and
$(i_3,s_2)$ and $(i_2,s_3)$.
Student $i_3$ appears in two cycles. So $i_3$ is assigned to her most
preferred school, $s_3$. Students $i_1$ and $i_2$ are assigned to the
schools they point to, $s_1$ and $s_2$, respectively.
After those students are removed and schools' capacities are
updated, only student $i_4$ and one seat $s_3$ remain. Student $i_4$
inherits the seat at $s_3$ and there is a self-cycle in Step 2 where
$(i_4,s_3)$ points to itself.

So, student $i_4$ cannot be assigned to any school in her interim safety-set
$\{s_1,s_2\}$.

\subsection{Application-Rejection mechanism}
\label{sec:appl-reject-mech}

Another $\Pi^{C}$-contestable mechanism is the Application-Rejection mechanism
described by \cite{chen2017chinese}. In broad strokes, this mechanism
relies on a multi-round algorithm, where each round is parametrized by
a \textit{permanency-execution period} $e$. Within each round $t$, a
deferred-acceptance algorithm is used by considering only the schools
that rank between $(t-1)e+1$ and $te$ in students' preferences. At the
end of each round, students who have been assigned are permanently
assigned and removed from the problem, and schools' capacities are
reduced by the number of students permanently assigned.

It is relatively intuitive to see this mechanism does not respect
safety-sets when $e$ is smaller than the number of schools. 
To see this, consider the  example depicted in
Table~\ref{tab:application-rejection}, where each school has only one
seat.  

\begin{table}[ht!]
  \centering
  \begin{tabular}{cccccccc}
$i_1$ & $i_2$ & $i_3$ & $i_4$ & \qquad & $s_1$ & $s_2$ & $s_3$\\
    \cline{1-4}\cline{6-8}
    $s_1$ & $s_1$ & $s_1$ & $s_3$ &  & $i_1$ & $i_1$ & $i_1$\\
$s_2$ & $s_2$ & $s_2$ & $s_1$ &  & $i_2$ & $i_2$ & $i_2$\\
$s_3$ & $s_3$ & $s_3$ & $s_2$ &  & $i_3$ & $i_3$ & $i_3$\\
 &  &  &  &  &   $i_4$ & $i_4$ & $i_4$
  \end{tabular}
  \caption{The Application-Rejection mechanism is $\Pi^{C}$-contestable}
  \label{tab:application-rejection}
\end{table}

We have $\Safe^I(\succ_{i_3},\{C\})=\{s_1,s_2,s_3\}$. Let $e=2$, so that the first
round only considers the first two choices of each student. 
In the first
round of the Application-Rejection algorithm, schools $s_1$, $s_2$,
and $s_3$ are filled with students $i_1$, $i_2$, and $i_4$,
respectively. So $i_3$ cannot be assigned to any school in her
safety-set.

\newpage

\bibliographystyle{apalike}
\bibliography{incontestable}